\documentclass[aps,prb,reprint,superscriptaddress,showpacs]{revtex4-1}
\usepackage{amsmath,amssymb}
\usepackage{amsbsy}
\usepackage{graphicx}

\newcommand{\vecM}{\mathbf{M}}
\newcommand{\VL}{\mathbf{V}_L}
\newcommand{\vecm}{\mathbf{m}}

\newcommand{\rot}[1]{\mathop{\mathrm{rot}}{#1}}

\renewcommand{\vec}[1]{\mathbf{#1}}
\newcommand{\mean}[1]{\left\langle #1 \right\rangle}
\newcommand{\abs}[1]{\left| #1 \right|}

\newcommand{\Meff}{\vec{M}_{\mathrm{eff}}}
\newcommand{\Heff}{\vec{H}_{\mathrm{eff}}}
\newcommand{\leff}{\lambda_{\mathrm{eff}}}
\newcommand{\Han}{H_{\mathrm{an}}}
\newcommand{\Vth}{V_{\mathrm{th}}}
\newcommand{\jth}{j_{\mathrm{th}}}

\begin{document}

\title{Magnon radiation by moving Abrikosov vortices in ferromagnetic superconductors and superconductor-ferromagnet multilayers}

\author{A. A. Bespalov}
\affiliation{Institute for Physics of Microstructures,
 Russian Academy of Sciences, GSP-105, 603950, Nizhny Novgorod, Russia}
\affiliation{Universit\'e Bordeaux, CNRS, LOMA, UMR 5798, F-33400 Talence, France}
\author{A. S. Mel'nikov}
\affiliation{Institute for Physics of Microstructures,
 Russian Academy of Sciences, GSP-105, 603950, Nizhny Novgorod, Russia}
\affiliation{Nizhny Novgorod State University, 22 Gagarin av., 603950, Nizhny Novgorod, Russia}
\author{A. I. Buzdin}
\affiliation{Universit\'e Bordeaux, CNRS, LOMA, UMR 5798, F-33400 Talence, France}

\begin{abstract}
In systems combining type-II superconductivity and magnetism the non-stationary magnetic field of moving Abrikosov vortices may excite spin waves, or magnons. This effect leads to the appearance of an additional damping force acting on the vortices. By solving the London and Landau-Lifshitz-Gilbert equations we calculate the magnetic moment induced force acting on vortices in ferromagnetic superconductors and superconductor/ferromagnet superlattices. If the vortices are driven by a dc force, magnon generation due to the Cherenkov resonance starts as the vortex velocity exceeds some threshold value. For an ideal vortex lattice this leads to an anisotropic contribution to the resistivity and to the appearance of resonance peaks on the current voltage characteristics. For a disordered vortex array the current will exhibit a step-like increase at some critical voltage. If the vortices are driven by an ac force with a frequency $\omega$, the interaction with magnetic moments will lead to a frequency-dependent magnetic contribution $\eta_M$ to the vortex viscosity. If $\omega$ is below the ferromagnetic resonance frequency $\omega_F$, vortices acquire additional inertia. For $\omega>\omega_F$ dissipation is enhanced due to magnon generation. The viscosity $\eta_M$ can be extracted from the surface impedance of the ferromagnetic superconductor. Estimates of the magnetic force acting on vortices for the U-based ferromagnetic superconductors and cuprate/manganite superlattices are given.
\end{abstract}

%\pacs{75.30.Ds, 74.25.Uv, 74.20.De}

\maketitle

\section{Introduction}

Within the last 13 years a number of fascinating compounds has been discovered, revealing the coexistence of ferromagnetism and superconductivity in the bulk.\cite{Saxena+2000,URhGe2000,UCoGe2007,Flouquet+2002} These compounds are U-based ferromagnets which become superconducting at temperatures $\sim$1K under applied pressure, or even at atmospheric pressure. Experimental investigation of magnetic properties of these materials in the superconducting state is hampered by the Meissner effect, making static measurements inefficient. However, important parameters can be extracted from dynamical measurements of the spin-wave (magnon) spectrum, which can be determined, e. g., by microwave probing \cite{Braude+Sonin2004,Braude2006} or using Abrikosov vortex motion.\cite{Bulaevskii+2005,Bulaevskii+2012}

A number of papers has been devoted to theoretical investigation of the magnon spectrum in magnetic superconductors.\cite{Buzdin84,Braude+Sonin2004,Braude+2005,Braude2006,Logoboy+2007,Ng+98,Bespalov+2013} Buzdin \cite{Buzdin84} determined the magnon spectrum in a superconducting antiferromagnet with an easy-axis anisotropy. Different types of spin waves in ferromagnetic superconductors in the Meissner state have been studied by Braude, Sonin and Logoboy,\cite{Braude+Sonin2004,Braude+2005,Braude2006,Logoboy+2007} including surface waves and domain wall waves.

Experimental measurements of the ac magnetic susceptibility of superconducting ferromagnets revealed that the screening of the magnetic field created by magnetic moments in these materials is incomplete.\cite{URhGe2000,UCoGe2007} This indicates that the superconducting transition in the U-compounds occurs in the spontaneous vortex state. Only two papers so far have addressed the influence of Abrikosov vortices on the magnon spectrum in ferromagnetic superconductors. In Ref. \onlinecite{Ng+98} coupled magnetic moment-vortex dynamics has been studied in the limit of long wavelength $\lambda_w \gg a$, where $a$ is the inter-vortex distance. Later,\cite{Bespalov+2013} this analysis has been extended to the case $\lambda_w \lesssim a$. It has been demonstrated that in the presence of a vortex lattice the magnon spectrum acquires a Bloch-like band structure.

To study the spin wave spectrum experimentally two simple procedures have been proposed. The first method consists in direct excitation of magnons by an electromagnetic wave incident at the sample.\cite{Braude+Sonin2004,Braude2006,Bespalov+2013} Then, information about the spin wave spectrum can be extracted from the frequency dependent surface impedance $Z$. Note that this procedure can be applied also to ordinary ferromagnets, but in all cases the high quality crystalline surface is required. The surface impedance has been calculated for a ferromagnetic superconductor in the Meissner state \cite{Braude+Sonin2004,Braude2006} and in the mixed state \cite{Bespalov+2013} for a static vortex lattice. 

The second method is based on the indirect magnon excitation: an external source of current sets in motion the Abrikosov vortices, which start to radiate magnons when the Cherenkov resonance condition is satisfied.\cite{Bulaevskii+2005,Bulaevskii+2012} Here, the current-voltage characteristics yield information about the magnon spectrum. Since this method involves Abrikosov vortices, it is specific for superconducting materials. Different phenomena arising from vortex-magnetic moment interaction in magnetic superconductors have been studied by Bulaevskii et al.\cite{Bulaevskii+2005,Bulaevskii+2011,Bulaevskii+2012,Bulaevskii+Lin2012,Lin+2012,Bulaevskii_polaronlike2012} In Refs. \onlinecite{Bulaevskii+2005,Bulaevskii+2011} the dissipation power due to magnon generation by a moving with a constant velocity vortex lattice in a superconducting antiferromagnet has been calculated. In Ref. \onlinecite{Bulaevskii+2012} this result has been generalized for the case of a vortex lattice driven by a superposition of ac and dc currents. In Refs. \onlinecite{Bulaevskii_polaronlike2012} and \onlinecite{Bulaevskii+Lin2012} a polaronic mechanism of self-induced vortex pinning in magnetic superconductors is discussed. The motion of the vortex lattice under the action of dc \cite{Bulaevskii_polaronlike2012} and ac currents \cite{Bulaevskii+Lin2012} has been studied. Finally, in Ref. \onlinecite{Lin+2012} it has
been predicted that the flux flow should lead to the creation of domain walls in systems with slow relaxation of the magnetic moments.

In the present paper, by solving the phenomenological London and Landau-Lifshitz equations, we analyze the problem of magnon generation by moving Abrikosov vortices in ferromagnetic superconductors and superconductor/ferromagnet (SF) multilayers. Theoretical investigation of the latter systems is relevant in view of the recent success in fabrication and characterization of cuprate-manganite superlattices.\cite{Multilayer,Hoppler+2009,Depleted} Also, recently an experimental study of the flux-flow resistivity in Nb/PdNi/Nb trilayers has been reported.\cite{Silva+2012} Our consideration of bulk ferromagnetic superconductors, on the other hand, is relevant to the U-based compounds mentioned above. In this aspect, the present work complements the preceding papers,\cite{Bulaevskii+2005,Bulaevskii+2011,Bulaevskii+2012,Bulaevskii+Lin2012,Bulaevskii_polaronlike2012,Lin+2012} which concetrated mainly on antiferromagnetic materials.
As we show, the presence of ferromagnetism introduces its own specifics, as the magnon spectrum in ferromagnets differs from the antiferromagnetic spectrum. Our results also include the comparison of the cases of a disordered and regular vortex lattice.

The outline of the paper is as follows. In Sec. \ref{sec:general} we give a model of the ferromagnetic superconductor and derive a general equation for the magnetic moment induced force $\vec{f}_M$ acting on vortices in ferromagnetic superconductors. In Sec. \ref{sec:constV} this force is calculated for a vortex lattice and disordered vortex array moving under the action of a dc transport current. Here, the differences in the dependence of $\vec{f}_M$ vs. vortex velocity for ferromagnetic and antiferromagnetic materials are discussed. Section \ref{sec:harmonic} is devoted to vortex motion under the action of an ac driving force. The magnetic contributions to the vortex viscosity and vortex mass are determined. In Sec. \ref{sec:Impedance} it is shown how the force $\vec{f}_M$ can be estimated experimentally by measuring the surface impedance. In Sec. \ref{sec:SFS} the generalization of our calculations for the SF multilayers is discussed. In the conclusion a summary of our results is given.

\section{The interaction force between vortices and magnetic moments: general equations}
\label{sec:general}

In the London approximation the free energy of the ferromagnetic superconductor in the mixed state can be taken in the form
\begin{eqnarray}
    & F = \int \left[ \frac{1}{8\pi \lambda^2} \left(\vec{A} + \frac{\Phi_0}{2\pi} \nabla \theta_S \right)^2 + \frac{(\rot{\vec{A}} - 4\pi \vec{M})^2}{8\pi} \right. & \nonumber \\
    & \left. + \frac{\alpha}{2} \left( \frac{\partial \mathbf{M}}{\partial x_i} \frac{\partial \mathbf{M}}{\partial x_i} \right) + \frac{K \mathbf{M}_{\bot}^2}{2} - \frac{\vec{B} \vec{H}_e}{4\pi} \right] d^3 \vec{r}. &
    \label{eq:F}
\end{eqnarray}
Here $\lambda$ is the London penetration depth, $\vec{A}$ is the vector potential, 
$\Phi_0$ is the flux quantum ($\Phi_0 = \pi \hbar c/\abs{e}>0$), $\theta_S$ is the superconducting order parameter phase, 
$\vec{M}$ is the magnetization, and $\alpha$ is a constant characterizing the exchange interaction. The U-based ferromagnetic superconductors,  listed in Table \ref{tab:parameters}, have a strong easy-axis magnetocrystalline anisotropy, which is accounted for by the term $K \vec{M}_{\bot}^2 / 2$, where $K$ is an anisotropy constant, $\vecM_{\bot} = \vecM - (\vec{e} \cdot \vecM) \vec{e}$, and $\vec{e}$ is a unit vector along the anisotropy axis. The term $\vec{B} \vec{H}_e/4\pi$ in Eq. \eqref{eq:F} accounts for a uniform external field $\vec{H}_e$. All terms in the right-hand side of Eq. \eqref{eq:F} are integrated over the whole space, except for the first term, containing $\lambda$, which is integrated over the sample volume. Certainly, outside the sample $\vec{M} = 0$. In the sample the magnetization modulus is constant.

\begin{table}[t]
    \centering
        \begin{tabular}{|c|c|c|c|}
        \hline
            Compound & $\mathrm{UGe}_2$ & UCoGe & URhGe \\
        \hline
            $\lambda$, nm  & 1000 & 1200 & 3450 \\ 
        \hline
            $L$, nm & 13,6 & 45 &  900 \\
        \hline
          $\Han$, T & $\sim 100$ & $\sim 10$ & $\sim 10$ \\
        \hline
         		$\mu_U$ & 1,4$\mu_B$ & 0,07$\mu_B$ & 0,3$\mu_B$ \\
        \hline
        		$\omega_F$, Hz & $\sim 10^{13}$ & $\sim 10^{10}$ & $\sim 10^{11}$ \\
        \hline
        		$\Vth$, cm/s & $\sim 10^7$ & $\sim 10^5$ & $\sim 10^7$ \\
        \hline
        		 $K = \Han/M$ & $\sim 10^4$ & $\sim 10^4$ & $\sim 10^3-10^4$ \\
        \hline
        \end{tabular}
    \caption{Parameters of some ferromagnetic superconductors. $L = \sqrt{\alpha/K}$ is the effective domain wall width, $\Han$ is the anisotropy field, $\mu_U$ is the magnetic moment per U atom, $\mu_B$ is the Bohr magneton, $\omega_F$ is the ferromagnetic resonance frequency, and $\Vth$ is the critical vortex velocity for magnon radiation (see Sec. \ref{sec:constV}). The data have been taken from Refs. \onlinecite{Dao+2011,Saxena+2000,Shick2002,Huy+2008}.}
    \label{tab:parameters}
\end{table}

First, we determine the equilibrium state by minimizing $F$ with respect to $\vec{M}$, and then with respect to $\vec{A}$ and $\theta_S$. We note that anisotropy field $H_{\mathrm{an}} = KM$ is typically very large (see Table \ref{tab:parameters}): it is comparable to or greater than the upper critical field. This means that the inequality $B \ll \Han$ holds for any internal field $B$ that does not suppress superconductivity. Then the transverse component of the magnetization $\vec{M}_{\bot}$ can be estimated as $M_{\bot} \lesssim B/K \ll M$. Since $K \gg 1$, in a zero approximation with respect to $K^{-1}$ we can neglect the transverse magnetization (even in the anisotropy energy, which appears to be proportional to $K^{-1}$). Then
\begin{eqnarray}
    & F(\vec{A},\theta_S)  \approx \int \left[ \frac{1}{8\pi \lambda^2} \left(\vec{A} + \frac{\Phi_0}{2\pi} \nabla \theta_S \right)^2 \right. & \nonumber \\
    & \left. + \frac{B^2}{8\pi} - \vec{B} \vec{M}_0 - \frac{\vec{B} \vec{H}_e}{4\pi}  + 2\pi M^2 \right] d^3 \vec{r},&
    \label{eq:F(A)}
\end{eqnarray}
where $\vec{M}_0 = M \vec{e}$. For an arbitrary shaped sample further minimization can not be performed analytically. Here we assume the ferromagnetic superconductor to be an ellipsoid. The results derived below should be also valid in the extreme cases of slabs and long cylinders. It is reasonable to assume that the average internal magnetic field $\vec{B}_0$ in an ellipsoidal sample will be uniform (compare with a dielectric ellipsoid in a uniform external field - see Ref. \onlinecite{Electrodynamics}). Denoting the superconductor volume as $V$, we can rewrite the free energy as
\begin{eqnarray}
    & F = V \left(f_S(B_0) - \vec{M}_0 \vec{B}_0 - \frac{\vec{B}_0 \vec{H}_e}{4\pi} \right) & \nonumber \\
    & + \int_{\vec{r} \notin V} \left[ \frac{B^2}{8\pi} - \frac{\vec{B} \vec{H}_e}{4\pi} \right] d^3 \vec{r} + \mathrm{const}, &
    \label{eq:F1}
\end{eqnarray}
where the constant does not depend on the magnetic induction $\vec{B}$, and $f_S$ is the free energy density of the vortex lattice:
\begin{equation}
    f_S(B_0) = \mean{\frac{1}{8\pi \lambda^2} \left(\vec{A} + \frac{\Phi_0}{2\pi} \nabla \theta_S \right)^2 + \frac{B^2}{8\pi}}.
    \label{eq:f_S}
\end{equation}
Averaging is performed over a volume that is much larger than the inter-vortex distance. The function $f_S (B_0)$ can be determined explicitly by solving the London equation \eqref{eq:London} with a given vortex lattice density, corresponding to the average field $B_0$. To transform the integral in Eq. \eqref{eq:F1} we introduce several quantities: the self-field of the sample $\vec{B}_S = \vec{B} - \vec{H}_e$, the magnetization $\vec{M}_S$ due to supercurrents, the effective full magnetization $\Meff = \vec{M}_0 + \vec{M}_S$, and the effective H-field $\Heff = \vec{B}_S - 4\pi \Meff$. Then the integral can be transformed as
\begin{eqnarray*}
    &  \int_{\vec{r} \notin V} \left[ \frac{B^2}{8\pi} - \frac{\vec{B} \vec{H}_e}{4\pi} \right] d^3 \vec{r} = \int_{\vec{r} \notin V} \frac{B_S^2}{8\pi} d^3\vec{r} - \int_{\vec{r} \notin V} \frac{H_e^2}{8\pi} d^3\vec{r} & \\
    & = \int \frac{\Heff^2}{8\pi} d^3\vec{r} - \int_{\vec{r} \in V} \frac{\Heff^2}{8\pi} d^3\vec{r} + \mathrm{const} & \\
    & =  \frac{V}{2} \Meff \left( \hat{N} - \frac{\hat{N}^2}{4\pi} \right) \Meff + \mathrm{const}. &
\end{eqnarray*}
Here $\hat{N}$ is the demagnetizing tensor, connecting the effective magnetization and effective field inside the sample: $\Heff = -\hat{N} \Meff$. Analytical and numerical values of $\hat{N}$ can be found in Ref. \onlinecite{Osborn45}. Finally, if we eliminate $\Meff$ using the relation
\[ \Meff = (4\pi - \hat{N})^{-1} (\vec{B}_0 - \vec{H}_e), \]
we obtain
\begin{eqnarray}
    & \frac{F}{V} = f_S(B_0) - \vec{M}_0 \vec{B}_0 - \frac{\vec{B}_0 \vec{H}_e}{4\pi} & \nonumber \\
    & + \frac{1}{8\pi} (\vec{B}_0 - \vec{H}_e) \hat{N} (4\pi - \hat{N})^{-1} (\vec{B}_0 - \vec{H}_e) + \mathrm{const}. &
    \label{eq:F2}
\end{eqnarray}
Here, the only variable is the internal field $\vec{B}_0$, which should be 
determined from the equation
\begin{equation}
	\frac{\partial F}{\partial \vec{B}_0} = 0.
	\label{eq:B0}
\end{equation}
Equations \eqref{eq:F2} and \eqref{eq:B0} completely define the equilibrium state of the ferromagnetic superconductor.

\begin{figure}[t]
	\centering
		\includegraphics[scale=0.3]{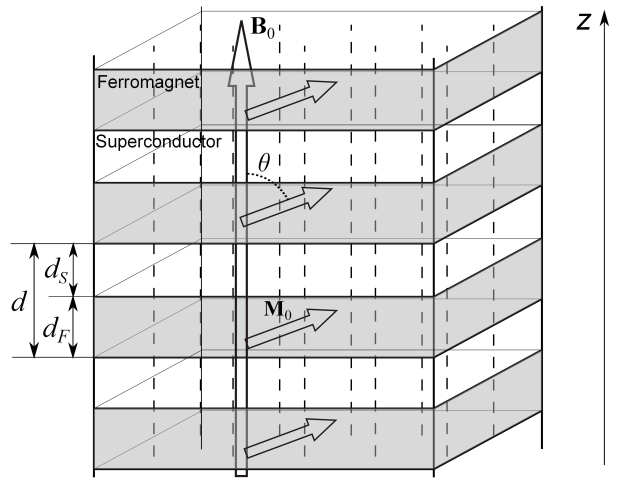}
		\caption{A scheme of the SF multilayer system. The dashed lines denote vortices.}
	\label{fig:SFS}
\end{figure}

Now we proceed from statics to coupled vortex and magnetization dynamics.
We focus on two systems, for which the derivation of the force acting on vortices is almost identical: a bulk ferromagnetic superconductor, and an SF multilayer (see Fig. \ref{fig:SFS}), where S is an ordinary type-II superconductor, and F is a ferromagnet with a strong easy-axis anisotropy: $K \gg 1$. For the multilayer system the same expression \eqref{eq:F} for the free energy is used with $\vec{M} = 0$ in the superconductor and $\lambda = \infty$ in the ferromagnet. We neglect the Josephson coupling between neighboring S layers. This is justified for $\gtrsim 10$ nm thick ferromagnets: in the case of an ordinary (non-triplet) proximity effect, superconducting correlations decay exponentially on a scale of several nanometers in the ferromagnet.\cite{Buzdin2005}

Let the vortices be aligned along the $z$-axis (which is perpendicular to the S/F interface
in the multilayer system). They may form a regular or disordered lattice.
When the vortices are set in motion by a dc or ac transport current, 
their time-dependent positions are given by the vector functions $\vec{R}_i(z,t)$, 
lying in the $xy$-plane, where $i=1..N_v$, and $N_v$ is the number of vortices.
In our calculations we will assume the vortices to be straight, i. e. $\vec{R}_i$ does not depend on $z$. 

As vortices move, the magnetic moments start to fluctuate. We describe the magnetization dynamics using the Landau-Lifshitz-Gilbert equation \cite{Gilbert}
\begin{equation}
    \frac{\partial \vecM}{\partial t} = \gamma \left( \vecM \times \frac{\delta F}{\delta \vecM} \right) + \frac{\nu}{M^2} \left( \vecM \times \frac{\partial \vecM}{\partial t} \right),
    \label{eq:LL}
\end{equation}
where $\gamma$ is the gyromagnetic ratio, $\nu$ is a dissipation constant, and the free energy $F$ is given by Eq. \eqref{eq:F}.

The force acting on a single vortex per unit length of the vortex equals
\begin{equation} 
  \vec{f}_i = - \frac{1}{L_v} \frac{\partial F}{\partial \vec{R}_i}.
  \label{eq:f_i}
\end{equation}
where $L_v$ is the vortex length. Averaging $\vec{f}_i$ over all vortices, we obtain the average force
\begin{equation}
    \vec{f} = - \frac{1}{L_v N_v}  \sum_i \frac{\partial F}{\partial \vec{R}_i},
    \label{eq:f1}
\end{equation}

When we considered the equilibrium state, the magnetization component perpendicular to the easy axis $\vec{e}$ has been neglected. Now we have to abandon this approximation, as it would lead to a vanishing force acting on the vortices from the side of the magnetic moments. We put $\vec{M} = \vec{M}_0 + \vec{m}$, where $\vec{m} \approx \vec{M}_{\bot}$, $\abs{\vec{m}} \ll M$, and linearize Eq. \eqref{eq:LL} with respect to $\vec{m}$:
\begin{equation}
    \frac{\partial \vecm}{\partial t} = -\gamma \vec{M}_0 \times \left(\alpha \nabla^2 \vecm - K\vecm + \vec{B} \right) + \frac{\nu}{M^2} \vec{M}_0 \times \frac{\partial \vecm}{\partial t}.
    \label{eq:LL_lin}
\end{equation}
From this equation it is evident that magnetization fluctuations are excited if the vortex field is not parallel to the magnetization easy-axis. In a ferromagnetic superconductor this may be achieved by applying an external field at an angle to the magnetization easy axis or by choosing an appropriate sample geometry (for example, an ellipsoidal sample with the magnetization directed along neither of the principal axes).

The magnetic induction inside the superconductor should be determined from the London equation
\begin{eqnarray}
    & \frac{\delta F}{\delta \vec{A}} = 0, \qquad \mbox{or} & \nonumber \\
    & -\nabla^2 \vec{B} + \frac{\vec{B}}{\lambda^2} = \frac{\Phi_0}{\lambda^2} \vec{z}_0 \sum_i \delta^{(2)} (\boldsymbol{\rho} - \vec{R}_i) + \frac{4\pi}{c} \rot{\rot{\vec{m}}}, \qquad &
    \label{eq:London}
\end{eqnarray}
where $\vec{z}_0$ is a unit vector along the $z$-axis. In the case of the multilayer system (see Fig. \ref{fig:SFS}), Maxwell's equations inside the F-layers read
\begin{equation}
  \rot{\vec{B}} = 4\pi \rot{\vec{M}}, \qquad \mathrm{div} \, \vec{B} = 0.
  \label{eq:Maxwell}
\end{equation}
On the SF-interface appropriate boundary conditions must be imposed:
\begin{eqnarray}
  & B_z \biggl|_F = B_z \biggl|_S, \qquad H_{x,y} \biggl|_F = H_{x,y} \biggl|_S & \label{eq:bound_BH} \\
  & \frac{\partial \vec{m}}{\partial z} \biggl|_F = 0. & \label{eq:bound_m}
\end{eqnarray}
The last condition follows directly from Eq. \eqref{eq:LL}, if no surface term is present in the free energy \eqref{eq:F}.

We present the magnetic field as the sum of the vortex field $\vec{h}$ and
the magnetization field $\vec{b}_M$ defined by
\begin{equation}
    -\nabla^2 \vec{h} + \frac{\vec{h}}{\lambda^2} = \frac{\Phi_0}{\lambda^2} \vec{z}_0 \sum_i \delta^{(2)} (\boldsymbol{\rho} - \vec{R}_i),
\label{eq:h}
\end{equation}
\begin{equation}
    -\nabla^2 \vec{b}_M + \frac{\vec{b}_M}{\lambda^2} = 4\pi \rot{\rot{\vec{m}}}
    \label{eq:b_M}
\end{equation}
inside the superconductor, and
\begin{equation}
  \rot{\vec{h}} = 0, \qquad \mathrm{div} \, \vec{h} = 0,
  \label{eq:h(F)}
\end{equation}
\begin{equation}
  \rot{\vec{b}_M} = 4\pi \rot{\vec{m}}, \qquad \mathrm{div} \, \vec{b}_M = 0
  \label{eq:b_M(F)}
\end{equation}
in the ferromagnetic layers.

In Eqs. \eqref{eq:London} - \eqref{eq:b_M(F)} we neglected the magnetic field induced by normal currents. These are given by $\vec{j} = \sigma \vec{E}$, where $\sigma$ is the normal conductivity, and $\vec{E}$ is the eletric field. We will first estimate the contribution of the normal currents flowing in the F-layers of the multilayer system to the magnetic field. Using both Maxwell's equations for $\rot{\vec{B}}$ and $\rot{\vec{E}}$, we obtain
\[ \rot{\rot{\vec{B}}} = \frac{4\pi \sigma_F}{c} \rot{\vec{E}} = -\frac{4\pi \sigma_F}{c^2} \frac{\partial \vec{B}}{\partial t}, \]
or 
\begin{equation}
	\frac{\partial^2 \vec{B}}{\partial z^2} + \nabla^2_{\boldsymbol{\rho}} \vec{B} -\frac{4\pi \sigma_F}{c^2} \frac{\partial \vec{B}}{\partial t} = 0,
	\label{eq:normal} 
\end{equation}
where $\sigma_F$ is the conductivity in the magnetic layers. Assuming
\[ \frac{\partial \vec{B}}{\partial t} \approx - (\VL \nabla_{\boldsymbol{\rho}}) \vec{B}, \]
where $\vec{V}_L$ is the flux velocity, we can see that the influence of the normal currents on the magnetic field is negligible, if the inequality
\[ \frac{4\pi \sigma_F}{c^2} lV_L \ll 1 \]
holds, where $l$ is the characteristic in-plane length scale of the problem. Similar arguments can be applied to the S-layers. Then, for the multilayer system we find the following constraint on the vortex velocity: 
\begin{equation}
  V_L \ll \frac{c^2}{4\pi \max(\sigma_n, \sigma_F) l},
  \label{eq:normal_mult}
\end{equation}
where $\sigma_n$ is the normal-state conductivity of the superconductor. In the case of a bulk ferromagnetic superconductor, we have to demand
\begin{equation}
	V_L \ll \frac{c^2}{4\pi \sigma_n l.}
	\label{eq:normal_FS}
\end{equation}
As we will see, the main length scales of the problem are the inter-vortex distance $a$ and the length $L = \sqrt{\alpha/K}$, which is of the order of the domain wall width of the ferromagnet (or ferromagnetic superconductor). Further on we assume that Eqs. \eqref{eq:normal_mult} and \eqref{eq:normal_FS} with $l=\min (a,L)$ are satisfied. Then, we may not take into account the normal currents in Eqs. \eqref{eq:London} - \eqref{eq:b_M(F)}.

In the free energy \eqref{eq:F} the interaction of the vortices 
with magnetization originates from the Zeeman-like term (a brief explanation is given in Appendix \ref{sec:Zeeman})
\[ F_Z = -\int \vec{M} \vec{h} d^3 \vec{r}. \]
Then, the magnetic moment induced force $\vec{f}_M$ acting on the vortices is
\begin{equation}
 \vec{f}_M = -\frac{1}{L_v N_v} \int m_z \nabla h_z d^3 \vec{r}.
    \label{eq:f1.5}
\end{equation}
Here it has been assumed that the perpendicular to the $z$-axis component of the field $\vec{h}$ is negligible. 
In the case of SF multilayers, this is true for a sufficiently small period of the structure.  To draw a parallel with preceding works, \cite{Bulaevskii+2005,Bulaevskii+2011,Bulaevskii+2012,Bulaevskii+Lin2012,Lin+2012,Bulaevskii_polaronlike2012} where the susceptibility formalism has been used, we note that $\vec{f}_M$ can be written in the form
\[ \vec{f}_M =  -\frac{1}{L_v N_v} \int (\hat{\chi}_{zz} h_z) \nabla h_z d^3 \vec{r}, \]
where $\hat{\chi}_{zz}$ is the susceptibility operator. If desired, the explicit form of $\hat{\chi}_{zz}$ can be easily derived from Eq. \eqref{eq:m_qz}, given below.

All further calculations in this section and Secs. \ref{sec:constV}, \ref{sec:harmonic} and \ref{sec:Impedance} are carried out for a ferromagnetic
superconductor. In Section \ref{sec:SFS} we discuss how our results can be extended to the case of the multilayer system. 

In the Fourier representation Eq. \eqref{eq:f1.5} reads
\begin{equation}
    \vec{f}_M = \frac{4\pi^2}{N_v}i \int \vec{q} m_{qz} h_{qz}^* d^2 \vec{q},
    \label{eq:f2}
\end{equation}
where for any function $X(\boldsymbol{\rho})$ its Fourier transform is defined as
\[  X_q = \frac{1}{(2\pi)^2} \int X(\boldsymbol{\rho}) e^{-i \vec{q} \boldsymbol{\rho}} d^2 \boldsymbol{\rho}. \]
By Fourier transforming Eqs. \eqref{eq:LL_lin}, \eqref{eq:h}, and \eqref{eq:b_M}, assuming that all quantities do not depend on $z$, we obtain
\begin{eqnarray}
    & \frac{\partial \vecm_q}{\partial t} = -\gamma \vec{M}_0 \times \left( - (K + \alpha q^2) \vecm_q + \vec{b}_{Mq} + \vec{h}_q \right) & \nonumber \\
    & + \frac{\nu}{M^2} \vec{M}_0 \times  \frac{\partial \vecm_q}{\partial t}, &
    \label{eq:LL_q}
\end{eqnarray}
\begin{equation}
    h_{qz} = \frac{\Phi_0}{4\pi^2 (1+\lambda^2 q^2)} \sum_i e^{-i\vec{q} \vec{R}_i (t)}.
    \label{eq:h_qz}
\end{equation}
\begin{equation}
    \vec{b}_{Mq} = -4\pi \frac{\vec{q} \times (\vec{q} \times \vec{m}_q)}{q^2 + \lambda^{-2}}.
    \label{eq:b_Mq}
\end{equation}
It can be seen that the absolute value of the term $\vec{b}_{Mq}$ in Eq. \eqref{eq:LL_q} 
is much smaller than $ \abs{K \vec{m}_q}$. Further on we will neglect the magnetization field $\vec{b}_{Mq}$.

Equation \eqref{eq:LL_q} is an inhomogeneous linear differential equation with constant coefficients with respect to $\vec{m}_q$. It can be solved using standard methods. We are interested in the $z$-component of the magnetization, which equals
\begin{eqnarray}
    & m_{qz} = \frac{\gamma M i}{2} \sin^2 \theta \int_{-\infty}^t h_{qz}(t') \left\{ \left( 1 + i\frac{\nu}{M} \right)^{-1} \right. & \nonumber \\
    & \times \exp{\left[ - \left(1 + i \frac{\nu}{M} \right)^{-1} i \omega(q) (t-t') \right]} & \nonumber \\
		& \left. - \left( 1 - i\frac{\nu}{M} \right)^{-1} \exp{\left[ \left(1 - i \frac{\nu}{M} \right)^{-1} i \omega(q) (t-t') \right]} \right\} dt', &
    \label{eq:mq'}
\end{eqnarray}
where $\theta$ is the angle between $\vec{e}$ and $\vec{z}_0$, and
\begin{equation}
	\omega(q) = \gamma M (K + \alpha q^2) = \omega_F (1+L^2 q^2)
	\label{eq:Spectrum}
\end{equation}
gives the magnon dispersion law in an ordinary ferromagnet, if the dipole-dipole interaction is not taken into account (see Ref. \onlinecite{Landau9}). Here, $\omega_F = \gamma M K$ is the ferromagnetic resonance frequency. In the small dissipation limit, $\nu \ll M$, we have
\begin{eqnarray}
  & m_{qz} = \frac{\gamma M i}{2} \sin^2 \theta \int_{-\infty}^t h_{qz}(t') \left\{ \exp{\left[ \left( - i - \frac{\nu}{M} \right)\omega(q)(t-t') \right]} \right. & \nonumber \\
  & \left. \times \left( 1 - i\frac{\nu}{M} \right) - \exp{\left[ \left( i - \frac{\nu}{M} \right) \omega(q) (t-t')\right]} \left( 1 + i\frac{\nu}{M} \right) \right\} dt'. & \nonumber \\
  \label{eq:m_qz}
\end{eqnarray}
Then, the force $\vec{f}_M$ takes the form
\begin{eqnarray}
    & \vec{f}_M =  \frac{2\pi^2 \gamma M}{N_v} \sin^2 \theta  \int d^2 \vec{q} \int_{-\infty}^t h_{qz}(t') h_{qz}^*(t) & \nonumber \\
    & \left\{ \exp{\left[ \left( i - \frac{\nu}{M}\right) \omega(q) (t-t') \right]} \left( 1 + i\frac{\nu}{M} \right) \right. & \nonumber \\
    &- \left. \exp{\left[ \left( - i - \frac{\nu}{M} \right) \omega(q) (t-t') \right]} \left( 1 - i\frac{\nu}{M} \right)\right\} \vec{q} dt'. &
    \label{eq:f5}
\end{eqnarray}

\section{Magnon radiation by vortices moving with a constant velocity}
\label{sec:constV}

Let us consider the motion of vortices under the action of a constant external force 
(e. g., spacially uniform and time-independent transport current).
Then the positions of individual vortices are given by
\begin{equation}
	\vec{R}_i (t) = \vec{R}_{i0} + \VL t + \Delta \vec{R}_i(t).
	\label{eq:R_constV}
\end{equation}
Here the vectors $\vec{R}_{i0}$ denote the vortex positions in a regular lattice, 
$\VL$ is the average flux velocity, and $\Delta \vec{R}_i(t)$ is responsible for fluctuations of
vortices due to interactions with pinning cites ($\mean{\Delta \vec{R}_i(t)} = 0$). 
It should be stressed here that we do not take into account the influence of pinning on the flux velocity. 
The effect that is important for us is the vortex lattice distortion caused by impurities,
which strongly influences the efficiency of magnon generation.

The product of magnetic fields under the integral in Eq. \eqref{eq:f5} is
\begin{equation}
	h_{qz}(t') h_{qz}^*(t) = \left( \frac{\Phi_0}{4\pi^2 (1+\lambda^2 q^2)} \right)^2 e^{i\vec{q} \VL (t-t')} {\cal K},
	\label{eq:hh}
\end{equation}
\begin{eqnarray*}
  & {\cal K} = \sum_{i,j} \exp{\left[ i\vec{q} (\vec{R}_{j0} -\vec{R}_{i0}) + i\vec{q}(\Delta \vec{R}_j(t) - \Delta \vec{R}_i(t')) \right]}. & %\\
\end{eqnarray*}

Below we consider the cases of a perfect vortex lattice and a  disordered vortex array.

\subsection{A perfect vortex lattice.}
\label{sub:VL_perfect}

The approximation used in this section is valid for sufficiently weak pinning, when we can put
\[ \mean{e^{i\vec{q}(\Delta \vec{R}_j(t) - \Delta \vec{R}_i(t'))}} \approx 1,\]
where the averaging is over $i$. To ensure the fulfillment of this condition it is sufficient to demand
\begin{equation}
	 \Delta R q \ll 1,
	 \label{eq:qRll1}
\end{equation}
where $\Delta R$ is the characteristic displacement of vortices from their positions in a perfect lattice. The inequality \eqref{eq:qRll1} must hold for all $\vec{q}$ giving a considerable contribution to the integral in Eq. \eqref{eq:f5}. In the end of Sec. \ref{sub:VL_disordered} it will be shown that this leads to the condition
\begin{equation}
	\Delta R \ll L.
	\label{eq:RllL}
\end{equation}
When \eqref{eq:RllL} holds, we have
\begin{equation}
	{\cal K} = \frac{4\pi^2 N_v B_0}{\Phi_0} \sum_{\vec{G}} \delta(\vec{q} - \vec{G}),
	\label{eq:K_perfect}
\end{equation}
where $\vec{G}$ are the vectors of the lattice, reciprocal to the vortex lattice. After integration over $\vec{q}$ and $t'$ the magnetic force takes the form
\begin{eqnarray*}
  & \vec{f}_M =  \Phi_0 B_0 \gamma M \sin^2 \theta \sum_{\vec G} \frac{\vec{G}}{(1+\lambda^2 G^2)^2} & \\
  & \times \frac{i\omega(G) + \frac{\nu}{M} \vec{G} \VL}{\omega^2(G) - (\VL \vec{G})^2 - 2i \frac{\nu}{M} \VL \vec{G} \omega(G)}.&
\end{eqnarray*}
When the terms corresponding to $\vec{G}$ and $-\vec{G}$ are combined, this can be written as
\begin{eqnarray}
  & \vec{f}_M = -\gamma \nu B_0 \Phi_0 \sin^2 \theta \sum_{\vec{G}} \frac{\vec{G} (\vec{G} \VL)}{(1+\lambda^2 G^2)^2} & \nonumber \\
  & \times \frac{(\vec{G} \VL)^2 + \omega^2(G)}{[\omega^2(G) - (\vec{G} \VL)^2]^2 + 4\frac{\nu^2}{M^2} (\vec{G} \VL)^2 \omega^2 (G)}, &
  \label{eq:f_VL_perfect}
\end{eqnarray}
where small terms of the order of $\nu/M$ in the numerator have been droped. From this it follows that the force has local maxima when for some $\vec{G} =\vec{G_0}$ the condition
\begin{equation}
  \omega(G_0) \approx \VL \vec{G_0}
  \label{eq:Cherenkov}
\end{equation}
is satisfied. This relation presents the well-known Cherenkov resonance condition. 
When Eq. \eqref{eq:Cherenkov} holds, magnons with the wave vector $\vec{G}_0$ are effectively generated.
When the vortex velocity is close to a resonance value, in the sum in Eq. \eqref{eq:f_VL_perfect} we can drop all terms except the two resonant terms corresponding to $\vec{G}_0$ and $-\vec{G}_0$. Then
\begin{eqnarray}
  & \vec{f}_M \approx -\gamma \nu B_0 \Phi_0 \sin^2 \theta \frac{\vec{G}_0}{(1+\lambda^2 G_0^2)^2} & \nonumber \\
  & \times \frac{\omega(G_0)}{(\omega(G_0) - \VL \vec{G}_0)^2 + \frac{\nu^2}{M^2} \omega^2(G_0).} &
  \label{eq:VL_peak}
\end{eqnarray}
It can be seen that the $\vec{f}_M$ vs. $V_L$ dependence for a given vortex velocity direction exhibits a
Lorentzian-like peak with the width 
\[ \Delta V_L = \frac{\nu}{M} \omega(G_0) \frac{V_L}{\vec{G}_0 \VL}. \]
The maximum value of $f_M$ is
\begin{equation}
  \abs{\vec{f}_M}_{\mathrm{max}} = \frac{\gamma M^2 B_0 \Phi_0 G_0 \sin^2 \theta}{(1+\lambda^2 G_0^2)^2 \nu \omega(G_0)}.
  \label{eq:f_Mmax}
\end{equation}
Another remarkeable feature is that the force is directed at some angle to the velocity of the vortices: $\vec{f}_M$ is parallel to $\vec{G}_0$, and not $\VL$. The angle between $\vec{f}_M$ and $\VL$ may range from $0^{\circ}$ to $90^{\circ}$. This effect also follows from Equation (3) in Ref. \onlinecite{Bulaevskii_polaronlike2012}, though the authors did not mention it, because it has been assumed that $\VL$ and $\vec{f}_M$ are always parallel.

Let us discuss how the Cherenkov resonances influence the current-voltage characteristics.
Abrikosov vortex motion in a superconductor is governed by the equation
\begin{equation}
  \frac{\Phi_0}{c} \vec{j} \times \vec{z}_0 = - \vec{f}
  \label{eq:Vortex_motion}
\end{equation}
The term in the left-hand side represents the Lorentz force with $\vec{j}$ being the macroscopic supercurrent density.
All other forces are represented by the term $\vec{f}$. We take into account two contributions to $\vec{f}$:
the viscous drag force $-\eta \VL$ and $\vec{f}_M$. 
Here, $\eta$ is the viscosity due to order parameter relaxation processes and normal current flowing through the
vortex core.\cite{Gor'kov+71} Taking the cross product of Eq. \eqref{eq:Vortex_motion} and $\vec{z}_0$, we
obtain the expression for the current
\begin{equation}
  \vec{j} = -\frac{c \eta}{\Phi_0} \VL \times \vec{z}_0 + \frac{c}{\Phi_0} \vec{f}_M(\VL)  \times \vec{z}_0.
  \label{eq:j}
\end{equation}
The relation between $\vec{j}$ and $\vec{E}$ is the established via
\begin{equation}
  \vec{E} = -\frac{1}{c} (\VL \times \vec{B}),
  \label{eq:Faraday}
\end{equation}
which follows from Faraday's law. According to Eq. \eqref{eq:j}, the vortex-magnetic moment interaction leads to an increase $\Delta \vec{j}$ of the current density at a given electric field $\vec{E}$:
\[ \Delta \vec{j} = \frac{c}{\Phi_0} \vec{f}_M \! \left( \frac{c}{B_0} \vec{E} \times \vec{z}_0 \right)  \times \vec{z}_0. \]
According to Eq. \eqref{eq:VL_peak}, near the Cherenkov resonance we have
\begin{eqnarray}
	& \Delta \vec{j} = \gamma \nu B_0 c \sin^2 \theta \frac{\vec{z}_0 \times \vec{G}_0}{(1+\lambda^2 G_0^2)^2} & \nonumber \\
	& \times \frac{\omega(G_0)}{\left[ \omega(G_0) - \frac{c}{B_0} (\vec{z}_0 \times \vec{G_0})\vec{E} \right]^2 + \frac{\nu^2}{M^2} \omega^2 (\vec{G}_0)}. &
	\label{eq:Delta_j_res}
\end{eqnarray}
This relation indicates that the I-V curve exhibits a series of peaks corresponding to the resonance electric fields given by 
\begin{equation}
	\omega(G) - \frac{c}{B_0} (\vec{z}_0 \times \vec{G})\vec{E} \approx 0
	\label{eq:E_resonance}
\end{equation}
Moreover, close to the resonance the additional current $\Delta \vec{j}$ is directed along the vector $\vec{z}_0 \times \vec{G}_0$ and not $\vec{E}$. The angle between $\Delta \vec{j}$ and $\vec{E}$ may range from $0^{\circ}$ to $90^{\circ}$. Thus, locally the resistance is anisotropic. 
Considering macroscopic ferrromagnetic superconductors and multilayer systems, care should be taken when applying Eq. \eqref{eq:Delta_j_res} to the whole sample: it is known that even a small concentration of pinning cites destroys the long-range order in the vortex lattice.\cite{Larkin70} In fact, vortex lattice domains are formed in large superconducting samples -- see Ref. \onlinecite{Domains} and references therein. The vortex nearest-neighbor directions are typically linked to crystal axes. Hence, in monocrystalline samples there are only few energetically favorable orientations of the vortex lattices. This fact allows us to put forward a qualitative argument. Let us denote as $\cal G$ the set of all reciprocal lattice vectors for all vortex lattice domains. Since there are only few possible orientations of the domains, the set $\cal G$ consists of isolated points. We claim that when the applied electric field satisfies Eq. \eqref{eq:E_resonance}
for some $\vec{G} \in \cal G$, the enhancement of the current should be observable. Hence, even if there are several vortex lattice domains, the peaks on the current-voltage characteristics are present. The measurement of the peak voltages at different applied magnetic fields allows to probe the magnon spectrum $\omega(q)$.

\subsection{A disordered vortex array}
\label{sub:VL_disordered}

In this section we analyze the opposite extreme case of chaotically placed vortices. This situation may be realized in weak magnetic fields, $B_0 \lesssim \Phi_0/\lambda^2$, when vortex-vortex interaction is weak and the lattice is easily destroyed by defects and thermal fluctuations.

We assume
\begin{equation}
	\mean{e^{i\vec{q}(\Delta \vec{R}_j(t) - \Delta \vec{R}_i(t'))}} \approx 0.
	\label{eq:VL_disorder_mean}
\end{equation}
where $q$ is not too small, and the averaging is over $i \neq j$. Note that the behavior of $\cal K$ at small $q$ almost does not influence the force $\vec{f}_M$, since $\cal K$ enters the integral in Eq. \eqref{eq:f5} with a factor $\vec{q}$. For $i = j$ we have
\[ \mean{e^{i\vec{q}(\Delta \vec{R}_i(t) - \Delta \vec{R}_i(t'))}} =1 \]
when $t = t'$, and
\[ \mean{e^{i\vec{q}(\Delta \vec{R}_i(t) - \Delta \vec{R}_i(t'))}} = \abs{\mean{e^{i\vec{q}\Delta \vec{R}_i(t)}}}^2 \ll 1 \]
when $\abs{t-t'} \to \infty$. To derive some qualitative results, we make the following assumption:
\[ \mean{e^{i\vec{q}(\Delta \vec{R}_i(t) - \Delta \vec{R}_i(t'))}} = e^{-\abs{t-t'}/\tau(q)}, \]
where the the time $\tau(q)$ is chosen so that $\mean{\abs{\vec{q}(\Delta \vec{R}_i(t) - \Delta \vec{R}_i(t'))}} \sim 1$ at $t-t' = \tau(q)$. Then
\begin{equation}
	{\cal K} = N_v e^{-\abs{t-t'}/\tau(q)},
	\label{eq:K_disorder}
\end{equation}
and after integration over $t'$ Eq. \eqref{eq:f5} yields
\begin{eqnarray}
	& \vec{f}_M = \frac{\gamma M \Phi_0^2 \sin^2 \theta}{8\pi^2} \int \frac{\vec{q} d^2 \vec{q}}{(1+\lambda^2 q^2)^2} \left[ \frac{1 + i\frac{\nu}{M}}{\tau^{-1}(q) + \frac{\nu}{M} \omega(q) - i\omega(q) - i\vec{q}\VL} \right.& \nonumber \\
	& \left. - \frac{1 - i\frac{\nu}{M}}{\tau^{-1}(q) + \frac{\nu}{M} \omega(q) + i\omega(q) - i\vec{q}\VL} \right]. &
	\label{eq:VL_disorder1}
\end{eqnarray}
It can be seen here that in the case of fast vortex fluctuations, $\tau^{-1}(q) \gg \omega(q)$, magnon generation is strongly suppressed, as the integral is proportional to $\tau(q)$. We will analyze in detail the opposite limiting case, $\tau^{-1}(q) \ll \omega(q)$. Then
\begin{eqnarray}
	& \vec{f}_M \approx \frac{\gamma M \Phi_0^2 \sin^2 \theta}{4\pi^2} \int \frac{\vec{q} d^2 \vec{q}}{(1+\lambda^2 q^2)^2} & \nonumber \\
	& \times \frac{i \omega(q)}{\omega^2(q) - (\vec{q} \VL)^2 - 2i(\vec{q} \VL)\tau_1^{-1}(q)}, &
	\label{eq:VL_disorder2}
\end{eqnarray}
where $\tau_1^{-1}(q) = \tau^{-1}(q) + \nu \omega(q)/M$, and in the numerator terms proportional to $\nu/M$ have been droped. The main contribution to the integral comes from $\vec{q}$ lying in the vicinity of two circles in the $\vec{q}$-plane, given by $\omega(q) = \pm \vec{q} \VL$ (this equation specifies the Cherenkov resonance condition). Near the circle $\omega(q) = \vec{q} \VL$ we make the following transformation:
\begin{eqnarray*}  
   & \omega^2(q) - (\vec{q} \VL)^2 - 2i(\vec{q} \VL)\tau_1^{-1}(q) & \\
   & \approx 2\omega(q) (\omega(q) - \vec{q} \VL -i\tau_1^{-1}(q)). &
\end{eqnarray*}
For the circle $\omega(q) = -\vec{q} \VL$ the transformations are analogous. Then
\begin{equation}
	\vec{f}_M \approx \frac{\gamma M \Phi_0^2 \sin^2 \theta}{4\pi^2} \!\! \int \!\! \frac{\vec{q} d^2 \vec{q}}{(1+\lambda^2 q^2)^2} \Re \frac{i}{\omega(q) - \vec{q} \VL -i\tau_1^{-1}(q)}.
	\label{eq:VL_disorder3}
\end{equation}
The last fraction in the right-hand side resembles the expression
\[ \Re \frac{i}{f(x) - i\epsilon}, \]
which reduces to $\delta(f(x))$ when $\epsilon \to +0$. Hence, the last factor in Eq. \eqref{eq:VL_disorder3} also can be replaced by a $\delta$-function, when $\tau_1^{-1}(q)$ is sufficiently small. To derive the limitation on $\tau_1^{-1}(q)$ we direct the $q_x$-axis along $\VL$ and rewrite the denominator of the large fraction in Eq. \eqref{eq:VL_disorder3} as follows:
\begin{eqnarray*} 
  & \omega_F (1 + L^2 q^2) - q_x V_L - i\tau^{-1}_1(q) & \\
  & = \omega_F \left( 1 - \frac{V_L^2}{\Vth^2} \right) - i \tau_1^{-1}(q) + \omega_F L^2 \!\! \left[ \left(q_x - \frac{V_L}{2L^2 \omega_F} \right)^2 \!\!\!\! + q_y^2 \right] \! , &
\end{eqnarray*}
where $\Vth = 2\omega_F L$. Now it is evident that the $\delta$-function can be introduced in Eq. \eqref{eq:VL_disorder3} when
\[ \tau^{-1}_1(q) \ll \omega_F \abs{\frac{V_L^2}{\Vth^2} - 1}. \]
Then
\begin{equation}
	\vec{f}_M \approx -\frac{\gamma M \Phi_0^2 \sin^2 \theta}{4\pi} \int \frac{\vec{q} d^2 \vec{q}}{(1+\lambda^2 q^2)^2} \delta(\omega(q) - \vec{q}\VL).
	\label{eq:VL_disorder4}
\end{equation}
Here, two points should be noted: (i) the expression for $\vec{f}_M$ does not depend on the dissipation rate and on the artificially introduced time $\tau(q)$; (ii) Equation \eqref{eq:VL_disorder4} can be derived from Eq. \eqref{eq:f_VL_perfect} in the limit of an extremely sparse vortex lattice, when summation can be replaced by integration.

Technical details of integration in Eq. \eqref{eq:VL_disorder4} are given in Appendix B. The final result is
\begin{equation}
	\vec{f}_M = -\frac{\gamma M \Phi_0^2 \sin^2 \theta}{8 \lambda^4 \omega_F^2} \left( 1 + \left( \frac{V_L}{\lambda \omega_F} \right)^2 \right)^{-3/2} \hspace{-0.5cm} \Theta (V_L - \Vth) \VL
	\label{eq:VL_disorder}
\end{equation}
for $\lambda \gg L$. Equation \eqref{eq:VL_disorder} asserts that the quantity $\Vth$ is the magnon generation threshold velocity. The maximal value of $f_M$ is reached at $V_L = \lambda \omega_F \sqrt{2} \gg \Vth$:
\begin{equation}
	\abs{\vec{f}_M}_{\mathrm{max}} = \frac{\Phi_0^2 \gamma M \sin^2 \theta}{8 \sqrt{2} \lambda^3 \omega_F 3^{3/2}}.
	\label{eq:fmax_rare}
\end{equation}

The influence of the magnetic force $\vec{f}_M$ on the current-voltage characteristics in general has been discussed in the previous section. According to Eqs. \eqref{eq:j} and \eqref{eq:Faraday}, at the electric field $E = \Vth B_0/c$ the average current density should exhibit a stepwise increase by
\[ \Delta j = \frac{c}{\Phi_0} f_M (\Vth) =  \frac{\gamma M \Phi_0 c \sin^2 \theta}{8 \lambda^4 \omega_F^2} \left( 1 + \left( \frac{\Vth}{\lambda \omega_F} \right)^2 \right)^{-3/2} \hspace{-0.5cm} \Vth. \]
The maximum enhancement of the current density due to vortex-magnetic moment interaction is reached at $E = \sqrt{2} \lambda \omega_F B_0/c$ and equals
\[ \Delta j_{\mathrm{max}} = \frac{\Phi_0 c \gamma M \sin^2 \theta}{8 \sqrt{2} \lambda^3 \omega_F 3^{3/2}}. \]

\begin{figure}[t]
	\centering
		\includegraphics[scale=0.3]{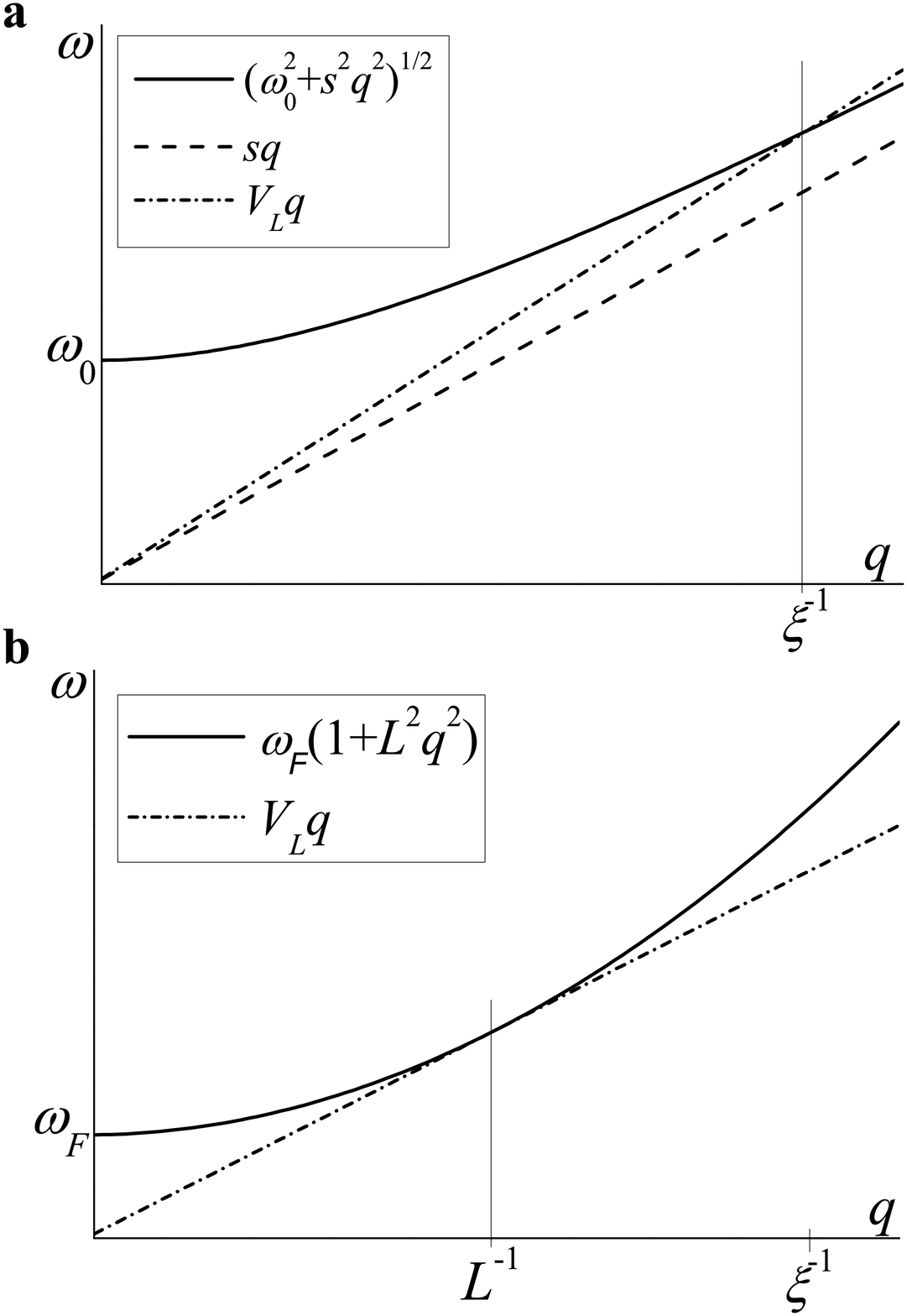}
	\caption{The magnon spectra in an (a) antiferromagnet and (b) ferromagnet. The dash-dotted line is given by $\omega = V_L q$, where $V_L$ is the vortex velocity at which magnon generation becomes efficient ($V_L =V_c$ for antiferromagnetic superconductors and $V_L = \Vth$ for ferromagnetic superconducors).}
	\label{fig:Spectra}
\end{figure}

In Ref. \onlinecite{Bulaevskii+2011} it has been predicted that in antiferromagnetic superconductors in the sparse lattice limit the current enhancement $\Delta j$ is proportional to $\sqrt{V_L - V_c}$ (at $V_L > V_c$), where $V_c$ is some critical velocity. This result is in contrast with ours: we found that $\Delta j \sim \Theta (V_L - \Vth)$ near the magnon generation threshold. This difference is due to different magnon spectra in ferromagnets and antiferromagnets -- see Fig. \ref{fig:Spectra}. In an antiferromagnet $\omega(q) = \sqrt{\omega_0^2 + s^2 q^2}$, where $\omega_0$ is a gap frequency, and $s$ is the short-wavelength magnon velocity. As the vortex velocity is increased, the resonance condition $\omega(q) = \VL \vec{q}$ is first satisfied at infinitely large $q$. However, at $q \gg \xi^{-1}$, where $\xi$ is the coherence length, the Fourier components $h_{qz}$ are exponentially small. Magnon generation becomes efficient at $q \sim \xi$, which is reached at a critical velocity that roughly equals $V_c = \sqrt{\omega_0^2 \xi^2 + s^2}$. In short, the generation threshold in antiferromagnetic superconductors corresponds to an intersection of the curves $\omega = \omega(q)$ and $\omega = V_L q$ at $q \sim \xi^{-1}$ (see Fig. \ref{fig:Spectra}a), yielding a $\Delta j \sim \sqrt{V_L - V_c}$ dependence. On the contrary, in ferromagnetic superconductors at $V_L = \Vth$ the curves $\omega = \omega(q)$ and $\omega = V_L q$ touch each other at $q = L^{-1}<\xi^{-1}$ (see Fig. \ref{fig:Spectra}b). This fact leads to a stepwise increase of the current at the threshold vortex velocity.

Finally, we need to make a remark concerning the condition \eqref{eq:qRll1}, providing that the ideal lattice approximation can be used. It follows from Fig. \ref{fig:Spectra}b that near the generation threshold magnons with wave numbers $q \approx L^{-1}$ are generated. This means that for $V_L \gtrsim \Vth$ the main contribution to the integral in Eq. \eqref{eq:f5} comes from $q \sim L^{-1}$. Thus, the condition \eqref{eq:RllL} should be imposed to ensure the applicability of the perfect lattice approximation.

\subsection{Estimates of the threshold vortex velocity in ferromagnetic superconductors and SF multilayers.}
\label{sub:U_check}

Let us check whether it is possible to observe the features connected with the Cherenkov resonances on the current-voltage characteristics of ferromagnetic superconductors and SF multilayers. To satisfy the condition \eqref{eq:Cherenkov} sufficiently large vortex velocities $V_L>\Vth$ are required. Estimates of the theshold velocity for known ferromagnetic superconductors are given in Table \ref{tab:parameters}. One can see that the values of $\Vth$ are very large. The question arises if such velocities are compatible with superconductivity in the U-based superconductors. To investigate this question we will estimate the supercurrent density $\jth$ which is sufficient to accelerate the vortices up to the velocity $\Vth$. Equation \eqref{eq:j} yields
\begin{equation}
	\jth \approx \frac{c\eta}{\Phi_0} \Vth
	\label{eq:j_th1}
\end{equation}
For the viscosity $\eta$ we use the Bardeen and Stephen expression \cite{Bardeen+65} (which is a good estimate for relatively slow processes)\cite{Kupriyanov+75}
\begin{equation}
	\eta = \Phi_0 H_{c2} \sigma_n/c^2,
	\label{eq:BS}
\end{equation}
where $H_{c2} = \Phi_0/(2\pi \xi^2)$ is the upper critical field. For the normal state conductivity we use Drude's estimate
\[ \sigma_n \sim \frac{e^2 n \ell}{m V_F}. \]
Here $n$ is the concentration of charge carriers, $m$ is their mass, $\ell$ is the mean free path, and $V_F$ is the Fermi velocity. Then
\begin{equation}
	\jth \sim \frac{e^2 n \ell H_{c2} \Vth}{mc V_F}.
	\label{eq:j_th2}
\end{equation}
This value should be compared with the depairing current density which is given within the BCS theory by
\[ j_{\mathrm{cr}} \sim en \frac{\Delta}{mV_F}, \]
where $\Delta$ is the superconducting gap. We demand $\jth \ll j_{\mathrm{cr}}$. Using the relation $\Delta \sim \hbar V_F/\xi$ (valid for clean superconductors) we can rewrite the inequality above as
\begin{equation}
	\Vth \ll \frac{\xi}{\ell} V_F.
	\label{eq:small_V_th}
\end{equation}
In the U-based compounds the coexistence of superconductivity and ferromagnetism appears in clean samples with $\ell \gtrsim \xi$. The Fermi velocities are of the order of $10^8 cm/s$ in $\mathrm{UGe_2}$ and $10^5 cm/s$ in UCoGe and URhGe -- see Refs. \onlinecite{Bauer+2001,Aoki+2011,Yelland+2011}. Thus, the inequality \eqref{eq:small_V_th} is satisfied in neither of these compounds, and our model breaks down at vortex velocities below $\Vth$. This is a consequence of the high magnetic anisotropy and large quasiparticle mass in the U-compounds.

The situation seems to be more optimistic in SF superlattices. Certainly, we should consider if Eq. \eqref{eq:BS} is valid for multilayers. A study of the vortex viscosity in superconductor/normal metal multilayers is presented in Refs. \onlinecite{Mel'nikovPRB96} and \onlinecite{Mel'nikovPRL96}. It has been shown the Bardeen-Stephen viscosity \eqref{eq:BS} may be significantly modified for vortices inclined with respect to the $z$-axis, or for strongly conducting normal metal layers. Still, in our case Eq. \eqref{eq:BS} is a good order-of magnitude estimate for $d_S \sim d_F$ and $\sigma_F \lesssim \sigma_n$, where $d_S$ and $d_F$ are the thicknesses of the superconducting and ferromagnetic layers (see Fig. \ref{fig:SFS}), respectively.

Recently, a number of experimental papers\cite{Multilayer,Hoppler+2009,Depleted} have reported successful fabrication of high-quality $\mathrm{YBa_2Cu_3O_7/La_{2/3}Ca_{1/3}MnO_3}$ superlattices. In Ref. \onlinecite{Mathur+2001} the value $\Han = 1200 Oe$ for $\mathrm{La_{0.7}Ca_{0.3}MnO_3}$ is given, though it is noted that the anisotropy is significantly influenced by strain. The measured domain wall width in the same compound, denoted as $\delta$ in Ref. \onlinecite{Lloyd+2001}, is 12 nm. Assuming $\gamma \sim \mu_B/\hbar$, where $\mu_B$ is the Bohr magneton, we obtain the following estimate for the vortex threshold velocity:
\begin{equation}
	\Vth = 2\gamma \Han L \sim 10^4 cm/s.
	\label{eq:V_LCMO}
\end{equation}
The Fermi velocity in $\mathrm{YBa_2Cu_3O_7}$ is of the order of or greater than $10^7 cm/s$.\cite{Hass+92} Thus, the condition \eqref{eq:small_V_th} can surely be satisfied in the cuprate/manganite superlattices.

\section{Magnon radiation by a harmonically oscillating vortex lattice}
\label{sec:harmonic}

As it has been shown in Sec. \ref{sub:U_check}, magnon generation in U-based ferromagnetic superconductors by a vortex array
moving with constant velocity seems problematic due to the extremely high required vortex velocities. In
this section we study a more feasible approach to magnon generation in magnetic superconductors, 
analyzing the case of a harmonic external current acting on the vortices. Experimentally, the oscillating current in the superconductor can be created using the microwave technique (for example, see Ref. \onlinecite{Silva+2012}). Then, the surface impedance yields information about the high-frequency properties of the sample -- see Sec. \ref{sec:Impedance}.
 
Subjected to the action of a harmonic force, in the linear regime the vortices oscillate harmonically:
\begin{equation}
  \vec{R}_i(t) = \vec{R}_{i0}' + \vec{R} e^{-i\omega t} + \vec{R}^* e^{-i\omega t}.
  \label{eq:R_oscillate}
\end{equation}
Here $\vec{R}_{i0}'$ are the equilibrium positions of the vortices, which are defined by vortex-vortex interaction
as well as pinning. The vectors $\vec{R}_{i0}'$ do not necessarily form a regular lattice, unlike
$\vec{R}_{i0}$. $\vec{R}$ is the amplitude of vortex oscillations. We will consider frequencies of the order of the
ferromagnetic resonance frequency in ferromagnetic superconductors, $\omega_F \sim 100 \mathrm{GHz}$.
This frequency is several orders of magnitude larger than the typical depinning frequency.\cite{Shapira+67}
This fact allows to neglect the influence of the pinning force on vortex motion and to assume that the
oscillation amplitudes of all vortices are equal to $\vec{R}$.

The product of the magnetic fields in Eq. \eqref{eq:f5} equals
\begin{eqnarray}
  & h_{qz}(t') h_{qz}^*(t) = \left( \frac{\Phi_0}{4\pi^2 (1+\lambda^2 q^2)} \right)^2 {\cal K'}
  e^{i\vec{q} (\vec{R}_i(t) - \vec{R}_i (t'))} & \nonumber \\
  & \approx \left( \frac{\Phi_0}{4\pi^2 (1+\lambda^2 q^2)} \right)^2 {\cal K'} 
  \left[ 1 + i\vec{q} (\vec{R}_i(t) - \vec{R}_i (t')) \right], &
  \label{eq:hh_harmonic}
\end{eqnarray}
where
\begin{equation}
  {\cal K'} = \sum_{i,j} e^{-i\vec{q} \vec{R}_{i0}' + i\vec{q} \vec{R}_{j0}'} 
  = N_v \mean{\sum_j e^{-i\vec{q} \vec{R}_{i0}' + i\vec{q} \vec{R}_{j0}'}}.
  \label{eq:K'}
\end{equation}
Here, the averaging is over $i$. The linear with respect to $\vec{R}$ contribution to the force takes the form
\begin{eqnarray}
  & \vec{f}_M =  \frac{\gamma M \Phi_0^2}{8 \pi^2 N_v} \sin^2 \theta  \int d^2 \vec{q} \int_{-\infty}^t 
  \frac{i {\cal K'} \vec{q} \vec{R}}{(1+ \lambda^2 q^2)^2} (e^{-i\omega t} - e^{-i\omega t'}) & \nonumber \\
  & \times \left\{ e^{\left[ i\omega(q) - \frac{\nu}{M} \omega(q) \right] (t-t')}
  - e^{\left[ - i\omega(q) - \frac{\nu}{M} \omega(q) \right] (t-t')} \right\} \vec{q} dt' + c. c.& \nonumber \\
  & \approx \frac{\gamma M \Phi_0^2}{4 \pi^2 N_v} \sin^2 \theta e^{-i\omega t} \int d^2 \vec{q} 
  \frac{{\cal K'}(q) \vec{q} \vec{R}}{(1+ \lambda^2 q^2)^2} & \nonumber \\
  & \times \left[ \frac{\omega(q)}{\omega^2(q) - \omega^2 - 2i \frac{\nu}{M} \omega \omega(q)} - \omega^{-1}(q) \right] \vec{q} + c.c.&
  \label{eq:f_harm}
\end{eqnarray}
Here $c. c.$ denotes the complex conjugate. Like before, we neglected small terms of the order of $\nu/M$.

To proceed further, the explicit form of ${\cal K'}(q)$ is required. 
Again, we will consider the cases of a pefect vortex lattice and a disordered array.

\subsection{A perfect vortex lattice.}

\label{sub:harmonic_lattice}

\begin{figure}[t]
	\centering
		\includegraphics[scale=0.3]{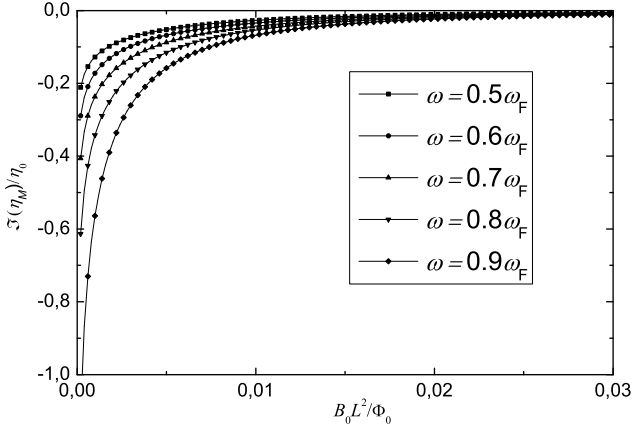}
	\caption{The $\Im(\eta_M)$ vs. magnetic field dependence at frequencies below the ferromagnetic resonance frequency for an ideal triangular vortex lattice (see Eq. \eqref{eq:eta_perfect2}). $\eta_0 = \gamma M \Phi_0^2 \sin^2 \theta/(2\lambda^4 \omega_F^2)$.}
	\label{fig:EtaBelow}
\end{figure}

Let us assume that pinning is sufficiently weak, so that
\begin{equation}
	q \Delta R \ll 1,
	\label{eq:perfect_qR}
\end{equation}
where $\Delta R \sim \abs{\vec{R}_{i0}' - \vec{R}_{i0}}$ is the characteristic deviation of the vortices from their positions in a perfect lattice. The inequality \eqref{eq:perfect_qR} should hold for all $\vec{q}$ giving a considerable contribution to the integral in Eq. \eqref{eq:f_harm}. The characteristic value of $q$ will be estimated below.

For $q \Delta R \ll 1$ we have
\begin{equation} 
   {\cal K'} = \frac{4\pi^2 N_v B_0}{\Phi_0} \sum_{\vec{G}} \delta(\vec{q} - \vec{G}).
   \label{eq:K'perfect}
\end{equation}
Substituting Eq. \eqref{eq:K'perfect} into Eq. \eqref{eq:f_harm}, assuming that the vortex lattice is either square or regular triangular, we obtain
\begin{equation}
  \vec{f}_M = i\omega \eta_M \vec{R} e^{-i\omega t} + c.c.,
  \label{eq:eta}
\end{equation}
\begin{eqnarray}
  & \eta_M = -\frac{i \gamma M \Phi_0 B_0}{2\omega} \sin^2 \theta \sum_{\vec{G}} 
  \frac{G^2}{(1+ \lambda^2 G^2)^2} & \nonumber \\ 
  & \times \left[ \frac{\omega(G)}{\omega^2(G) - \omega^2 - 2i \frac{\nu}{M} \omega \omega(G)} - \omega^{-1}(G) \right]. & 
  \label{eq:eta_perfect1}
\end{eqnarray}
Here, we have introduced the complex quantity $\eta_M$, playing the role of a generalized vortex viscosity. Indeed, when $\eta_M$ is purely real, the magnetic force is simply $\vec{f}_M = -\eta_M d \vec{R}_i/dt$. In our system there is a phase shift between the vortex velocity and $\vec{f}_M$, and the more general expression \eqref{eq:eta} is valid. Further on we will call $\eta_M$ the magnetic viscosity.

The ideal vortex lattice is likely to form when vortex-vortex interaction is sufficiently strong, or the
inter-vortex distance is suffiently small. Let this distance be much smaller than the London penetration depth, which means
$B_0 \gg \Phi_0/\lambda^2$. Then $\lambda G \gg 1$ for all $\vec{G} \neq 0$, and
\begin{eqnarray}
  & \eta_M \approx -\frac{i \gamma M \Phi_0 B_0 \omega}{2\lambda^4} \sin^2 \theta \sum_{\vec{G} \neq 0} 
  G^{-2} \omega^{-1}(G) & \nonumber \\
  & \times \left[ \omega^2(G) - \omega^2 - 2i \frac{\nu}{M} \omega \omega(G) \right]^{-1}. &
  \label{eq:eta_perfect2}
\end{eqnarray}

Now we consider the behavior of $\eta_M$ in different frequency ranges.
First, let the frequency be below the ferromagnetic resonance frequency ($\omega < \omega_M$). Then magnon
generation is inefficient. However, if we put $\nu=0$, the force $\vec{f}_M$ will not vanish below the generation threshold, unlike in the case of constant vortex velocity. Instead, the magnetic viscosity will be purely imaginary,
signifying that there are no magnetic losses. In Fig. \ref{fig:EtaBelow} we plot the imaginary part of $\eta$ vs. magnetic field
$B_0$ dependencies for different frequencies (below $\omega_F$) and for a fixed angle $\theta$.

Returning to the condition \eqref{eq:perfect_qR}, one can see that for $\omega<\omega_F$ the characterisitic values of $G$ in Eq. \eqref{eq:eta_perfect2} are of the order of $L^{-1}$. Hence, $\Delta R \ll L$ is required for Eq. \eqref{eq:eta_perfect2} to be valid.

\begin{figure}[t]
	\centering
		\includegraphics[scale=0.3]{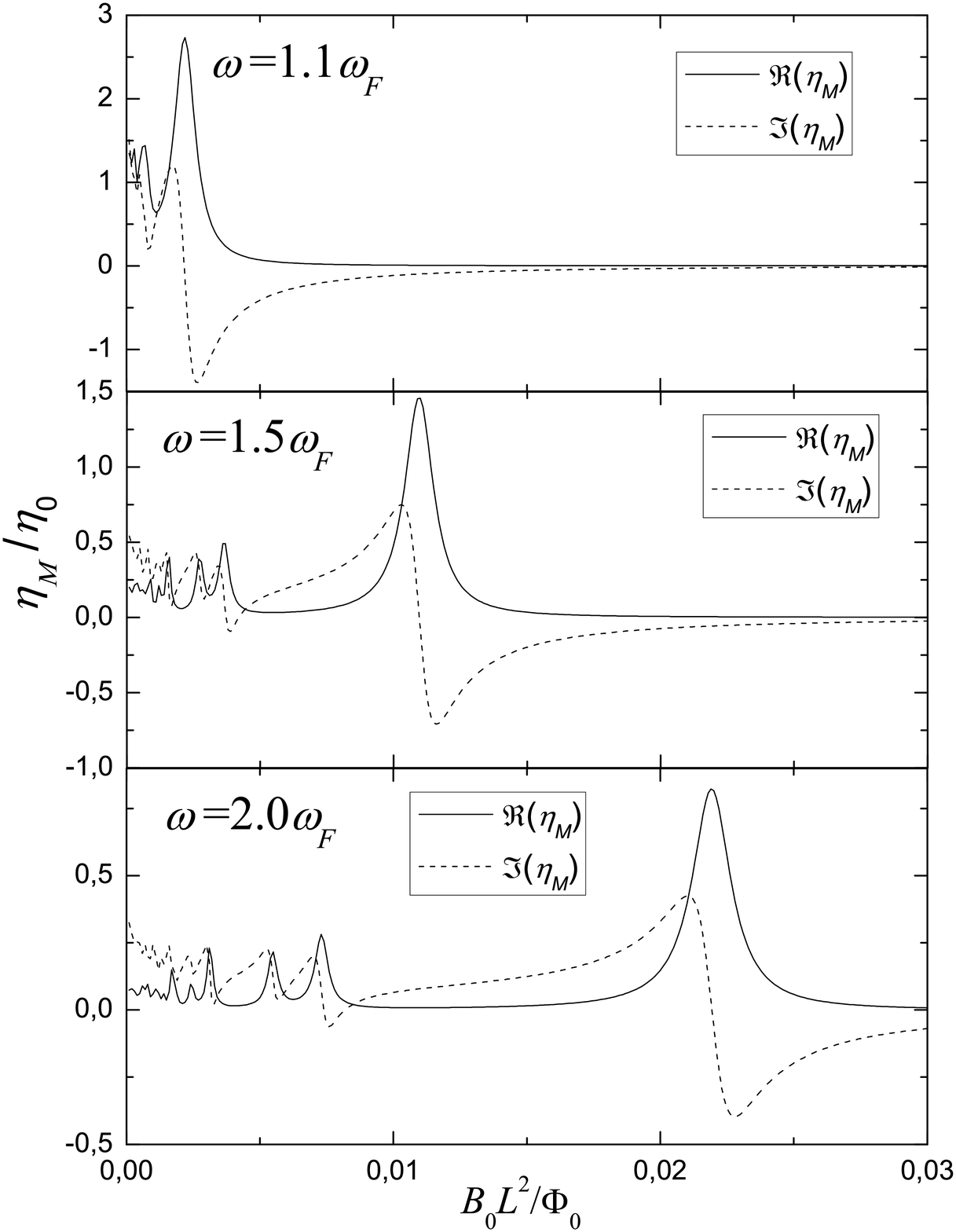}
	\caption{The $\eta_M$ vs. magnetic field dependencies for frequencies above the ferromagnetic resonance frequency (see Eq. \eqref{eq:eta_perfect2}). $\eta_0 = \gamma M \Phi_0^2 \sin^2 \theta/(2\lambda^4 \omega_F^2)$. The vortices form an ideal triangular lattice.}
	\label{fig:EtaAbove}
\end{figure}

At frequencies above the ferromagnetic resonance frequency magnetic dissipation can not be neglected, and the real 
part of $\eta_M$ becomes significant. 
In Fig. \ref{fig:EtaAbove} we plot the $\eta_M$ vs. $B_0$ dependencies for different frequencies and for a fixed angle $\theta$ and dissipation rate $\nu/M = 0.02$. 
The graphs exhibit a sequence of Lorentzian-like ($\Re(\eta_M)$) and N-shaped ($\Im(\eta_M)$) features, located at
some resonant field values, $B_R$, which are determined from the relation
\begin{equation}
  \omega(G) = \omega.
  \label{eq:res_harmonic}
\end{equation}
For small fields these features may overlap, but the resonance corresponding to the highest field remains well
distinguishable. For a triangular vortex lattice the largest resonance field equals
\[ B_R = \frac{\sqrt{3}}{8\pi^2} \frac{\Phi_0}{L^2} \left( \frac{\omega}{\omega_F} - 1 \right), \]
and for a square lattice
\[ B_R = \frac{1}{4\pi^2} \frac{\Phi_0}{L^2} \left( \frac{\omega}{\omega_F} - 1 \right). \]

Solving Eq. \eqref{eq:res_harmonic} with respect to $G$, we obtain
\begin{equation}
	G = L^{-1} \sqrt{\frac{\omega - \omega_F}{\omega_F}}.
	\label{eq:G_harmonic}
\end{equation}
Hence, the peaks on the $\eta_M$ vs. $B_0$ dependences must be observable if the characteristic deviation $\Delta R$ of the vortices from their positions in an ideal lattice satisfies
\begin{equation}
	\Delta R \ll G^{-1} = L \sqrt{\frac{\omega_F}{\omega - \omega_F}}.
	\label{eq:DeltaR_harmonic}
\end{equation}
Note that for frequencies close to $\omega_F$ this condition is weaker than $\Delta R \ll L$.

We conclude this section by giving a numeric estimate of the magnetic viscosity. When the resonance condition \eqref{eq:res_harmonic} is satisfied, we obtain from Eq. \eqref{eq:eta_perfect2}
\[ \eta_M \sim \frac{\gamma M \Phi_0 B_0}{\lambda^4 G^2 \omega^2} \frac{M}{\nu}. \]
Since $B_0 G^{-2} \sim \Phi_0$, and the lowest allowable value of $\omega$ is $\omega_F = \gamma M K$, we have
\begin{equation}
	\eta_M \lesssim \frac{\Phi_0^2}{K \lambda^4 \omega_F} \frac{M}{\nu}.
	\label{eq:etaM_estim1}
\end{equation}
Then, according to Eq. \eqref{eq:BS}, the ratio of $\eta_M$ to $\eta$ is
\begin{equation}
	\eta_M / \eta \lesssim \frac{M}{\nu} \frac{\xi^2 c^2}{K \lambda^4 \omega_F \sigma_n}.
	\label{eq:eta_estim2}
\end{equation}
We will make the numeric estimate for UCoGe, the ferromagnetic superconductor with the lowest ferromagnetic resonance frequency. In Ref. \onlinecite{UCoGe2007} we find the value $12\mu \Omega {\mathrm cm}$ for the normal resistivity, and the maximal value 200\AA \, for the coherence length. Using Table \ref{tab:parameters}, we obtain
\begin{equation}
	\eta_M/\eta \sim \frac{M}{\nu} 3 \times 10^{-5}.
	\label{eq:eta_estim3}
\end{equation}
Data on the ratio $M/\nu$ are not available yet. The small factor $10^{-5}$ in Eq. \eqref{eq:eta_estim3} appears due to the large magnetocrystalline anisotropy of UCoGe: it can be seen from Eq. \eqref{eq:eta_estim2} that $\eta_M/\eta$ is proportional to $K^{-2}$, since $\omega_F = \gamma MK$. Hence, to increase the ratio of $\eta_M$ to $\eta$, compounds (or multilayer systems) with a lower anisotropy are preferable.

\subsection{A disordered vortex array.}
\label{sub:harmonic_disorder}

\begin{figure}[t]
	\centering
		\includegraphics[scale=0.3]{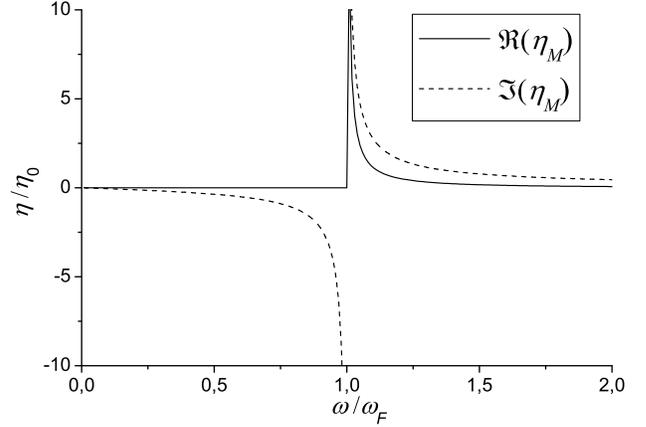}
	\caption{The frequency dependence of the magnetic viscosity, $\eta_M$, for a disordered vortex array  -- see Eq. \eqref{eq:harm_disorder3}. The value $\ln(\lambda/L) = 4.3$ of $\mathrm{UGe_2}$ has been used. $\eta_0 = \gamma M \Phi_0^2 \sin^2 \theta/(2\lambda^4 \omega_F^2)$.}
	\label{fig:HarmDisordered}
\end{figure}

Now let us assume that due to relatively strong pinning the vortices are placed chaotically, i. e., there is no correlation between
$\vec{R}'_{i0}$ and $\vec{R}'_{j0}$ for $i \neq j$. Then
\begin{eqnarray}
  & {\cal K'} = N_v \left( \int \frac{B_0}{\Phi_0} 
  e^{-i\vec{q} \vec{R}_{i0}' + i\vec{q} \vec{R}_{j0}'} d^2 \vec{R}'_{j0} + 1 \right) & \nonumber \\
  & = N_v \left( 4\pi^2 \frac{B_0}{\Phi_0} \delta (\vec{q}) + 1 \right). &
  \label{eq:K'_disorder}
\end{eqnarray}
In a real vortex lattice there is a short-range order, which leads to the smearing of the delta-function on a scale
of the order of the inverse inter-vortex distance. 
However, the behavior of ${\cal K'}$ at small $q$ is not important, since ${\cal K'}$ enters the integral in
Eq. \eqref{eq:f_harm} with a factor $q^2$. For clarity, we stress here that the product $h_{qz}(t') h_{qz}^*(t)$ does not decay with increasing $t-t'$, unlike in the case of a constant driving force -- see Eqs. \eqref{eq:hh} and \eqref{eq:K_disorder}. This is explained by the fact that the vortices oscillate close to their equilibrium positions and do not travel from one pinning cite to another. Thus, the positions $\vec{R}_i(t)$ and $\vec{R}_i(t')$ of a single vortex are always well correlated, corresponding to an infinite correlation time $\tau(q)$.

With $\cal K'$ given by Eq. \eqref{eq:K'_disorder} the magnetic viscosity takes the form
\begin{eqnarray}
  & \eta_M = -\frac{i\gamma M \Phi_0^2}{4\pi \omega} \sin^2 \theta \int_0^{\infty} \frac{q^3 dq}{(1+q^2 \lambda^2)^2} & \nonumber \\
  & \times \left[ \frac{\omega(q)}{\omega^2(q) - \omega^2 - i\epsilon} - \omega^{-1}(q) \right]. &
  \label{eq:harm_disorder1}
\end{eqnarray}
Here, like in Sec. \ref{sub:VL_disordered}, we assume that the imaginary term $-i\epsilon$ ($\epsilon>0$) in the denominator is an infinitesimal.
To simplify the expression in the right-hand side of Eq. \eqref{eq:harm_disorder1}, we note that
the contribution to the integral from small $q$ ($q \lesssim \lambda^{-1}$) can be neglected in the $\lambda \gg L$
limit. Then we can put $1+\lambda^2 q^2 \approx \lambda^2 q^2$, and cut the integral off at $q=\lambda^{-1}$:
\begin{eqnarray}
  & \eta_M = -\frac{i\gamma M \Phi_0^2}{4\pi \omega \lambda^4} \int_{\lambda^{-1}}^{\infty} \frac{dq}{q} & \nonumber \\
  & \times \left[ \frac{\omega(q)}{(\omega + \omega(q))(\omega(q) - \omega -i\epsilon)} - \omega^{-1}(q) \right]. &
  \label{eq:harm_disorder2}
\end{eqnarray}
Further integration should not present difficulties. For $\lambda \gg L$ we obtain
\begin{eqnarray}
  & \eta_M = \frac{\gamma M \Phi_0^2 \sin^2 \theta}{8\pi \omega \omega_F \lambda^4} \left\{ 
  \frac{\pi \omega_F}{2(\omega - \omega_F)} \Theta(\omega - \omega_F) - i \left[ \frac{2\omega^2}{\omega_F^2 - \omega^2} \ln \frac{\lambda}{L} \right. \right. & \nonumber \\
  & \left.  \left. +  \frac{\omega_F}{2(\omega + \omega_F)} \ln \frac{\omega + \omega_F}{\omega_F}
  + \frac{\omega_F}{2(\omega_F - \omega)} \ln \abs{\frac{\omega_F - \omega}{\omega_F}} \right] \right\}. &
  \label{eq:harm_disorder3}
\end{eqnarray}
Like in the previous section, below the ferromagnetic resonance frequency the magnetic viscosity is purely imaginary.
However, unlike in the case of a perfect fortex lattice, now the viscosity does not depend on the magnetic field.
It should be also noted that in the limit $B_0 \to 0$ Eq. \eqref{eq:eta_perfect1} after summation transforms into 
\eqref{eq:harm_disorder3}, i. e., the cases of isolated vortices and chaotically placed vortices are equivalent, like in Sec. \ref{sec:constV}. The $\eta_M$ vs. $\omega$ dependence is depicted in Fig. \ref{fig:HarmDisordered}.

\subsection{Vortex mass.}
\label{sub:mass}

As we have seen, at $\omega<\omega_F$ the magnetic viscosity is imaginary. 
Moreover, at $\omega \ll \omega_F$ the viscosity is proportional to $\omega$.
This signifies that the vortex can be ascribed a mass per unit length, $M_v$, so that the equation of motion becomes
\begin{equation}
  M_v \frac{d^2 \vec{R}_i}{dt^2} = \vec{f}_{\mathrm{ext}},
  \label{eq:with_mass}
\end{equation}
where $\vec{f}_{\mathrm{ext}}$ includes all forces, except for the force $\vec{f}_M$.
The mass is defined by
\begin{equation}
  M_v = \frac{i\eta_M}{\omega} \biggl|_{\omega=0}.
  \label{eq:mass_definition}
\end{equation}

Before we give explicit expressions for $M_v$, we should comment on the connection between the vortex mass enhancement and the self-induced polaronic pinning mechanism, studied in Refs. \onlinecite{Bulaevskii_polaronlike2012} and \onlinecite{Bulaevskii+Lin2012}. In the mentioned papers it has been assumed that the magnetization dynamics is purely dissipative, i. e., $\nu/M \gg 1$, which is in contrast to our case. In fact, $\nu/M \gg 1$ is a necessary condition for the formation of polaronlike vortices. Thus, the polaronic pinning mechanism contributes rather to the real part of $\eta_M$ than to its imaginary part, an it is not related to the vortex mass enhancement discussed here.

Using Eq. \eqref{eq:eta_perfect2}, we find that the magnetic contribution to the vortex mass for a perfect lattice is
\begin{equation}
  M_v = \frac{\gamma M \Phi_0 B_0}{2\lambda^4} \sin^2 \theta \sum_{\vec{G} \neq 0} G^{-2} \omega^{-3}(G)
  \label{eq:M_perfect}
\end{equation}
when $B_0 \gg \Phi_0/\lambda^2$. For a disordered array we obtain from Eq. \eqref{eq:harm_disorder1}
\begin{eqnarray}
  & M_v = \frac{\gamma M \Phi_0^2}{4\pi \omega_F^3} \sin^2 \theta \int_0^{\infty} \frac{q^3 dq}{(1+q^2\lambda^2)^2 (1+L^2 q^2)^3} & \nonumber \\
  & \approx \frac{\gamma M \Phi_0^2 \sin^2 \theta}{16\pi \omega_F^3 \lambda^4} 
  \left(4\ln \frac{\lambda}{L} - 5 \right) \qquad (\lambda \gg L). &
  \label{eq:M_disorder}
\end{eqnarray}
Let us estimate the characteristic magnetic contribution $M_v$ to the vortex mass and compare it with the
electronic contribution (see, for example, Ref. \onlinecite{Vortex_review}), which is present in any superconductor:
\begin{equation}
  M_e = \frac{2}{\pi^3} \frac{m^2 V_F}{\hbar}.
  \label{eq:M_e}
\end{equation}
We give estimates for the ferromagnetic superconductor URhGe. The values of $\omega_F = \gamma M K$ and $\lambda$ can be found in Table \ref{tab:parameters}. The electron mass and Fermi velocity for one of the
Fermi surface pockets of URhGe have been measured in Ref. \onlinecite{Yelland+2011}. The values given there are
$m = 22m_e$ and $V_F = 4.4 \times 10^5 cm/s$, where $m_e$ is the free electron mass. Then
\[ M_v \sim \frac{\gamma M \Phi_0^2}{16 \pi \omega_F^3 \lambda^4} \approx 10^{-24}g/cm, \qquad
M_e \sim 10^{-20} g/cm. \]
It can be seen that the magnetic contribution to the vortex mass is negligible for URhGe. Estimates for $\mathrm{UGe_2}$ and UCoGe yield the same result. This happens due to the very large ferromagnetic resonance frequency $\omega_F$ in these compounds: note that the right-hand side of Eq. \eqref{eq:M_disorder} contains $\omega_F^{-3}$. The situation is the same as for the magnetic viscosity -- see Eqs. \eqref{eq:eta_estim2} and \eqref{eq:eta_estim3}. Thus, the magnetic mass $M_v$ should be detectable in materials with a smaller ferromagnetic resonance frequency.

\section{Discussion of the magnetic viscosity measurement}
\label{sec:Impedance}

\begin{figure}[t]
	\centering
		\includegraphics[scale=0.3]{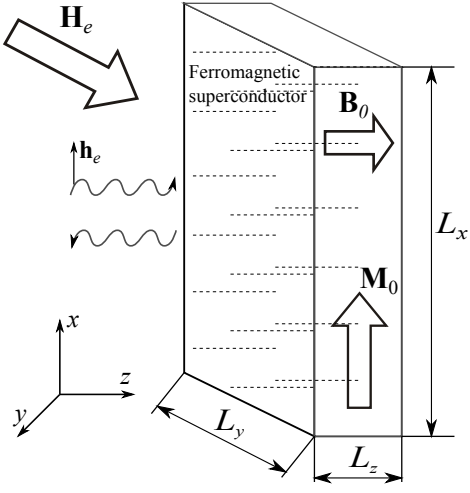}
	\caption{The geometry for the measurement of the surface impedance of a ferromagnetic superconductor. The dashed lines denote vortices.}
	\label{fig:Both}
\end{figure}

A simple experimental method to study vortex dynamics in type-II superconductors consists in the measurement of the surface impedance. A possible geometry for such experiment is depicted in Fig. \ref{fig:Both}. We consider the simplest situation, when the vortices are perpendicular to the sample surface, and the probing electromagnetic wave with the amplitude $\vec{h}_e$ is normally incident on this surface. Then, for a non-magnetic superconductor ($\vec{M}_0 = 0$) theory\cite{Gor'kov+71,Sonin+Traito94} predicts that in a wide range of parameters the surface impedance $Z(\omega)$ equals
\begin{equation}
  Z(\omega) = \left( \frac{-i\omega \mu \rho_f}{4\pi} \right)^{1/2},
  \label{eq:Z}
\end{equation}
where $\mu$ is the static differential magnetic permeability,
\[ \mu = \frac{dB_{0z}}{dH_{ez}}, \]
and $\rho_f$ is the flux-flow resistance,
\[ \rho_f = \frac{B \Phi_0}{c^2 \eta}. \]
Thus, the experimental value of the surface impedance provides information about the viscosity coefficient $\eta$.

We will prove that for a ferromagnetic superconductor a range of parameters exists, where Eq. \eqref{eq:Z} can be applied, if the magnetic viscosity is taken into account: $\eta$ should be replaced by $\eta+ \eta_M$. 

First, we outline the applicability conditions of Eq. \eqref{eq:Z} for an ordinary superconductor. Within the continuous medium approximation used in Ref. \onlinecite{Sonin+Traito94} an alternating external field $\vec{h}_{e}$ excites a long-wavelength and a short-wavelength mode in the superconductor. For convenience, we will call these modes type-1 and type-2, and denote the $z$-projections of their wave vectors as $k_1$ and $k_2$, respectively. These quantities are explicitly defined by Equation (24) in Ref. \onlinecite{Sonin+Traito94}. To use the simple expression \eqref{eq:Z} for the impedance, three conditions must be fulfilled: (i) $\abs{k_1} \lambda \ll 1$, (ii) $\abs{k_1} \ll \abs{k_2}$, and (iii) $\abs{k_2} L_z \gg 1$ ($L_z$ is the sample thickness -- see Fig. \ref{fig:Both}). According to Ref. \onlinecite{Sonin+Traito94}, the conditions (i) and (ii) are satisfied, if
\begin{equation}
  \omega \ll \omega_C = \frac{\Phi_0 C^*_{44}}{B_0 \lambda^2 \eta},
  \label{eq:w_ll_wC}
\end{equation} 
where $C^*_{44}$ is an elastic modulus of the vortex lattice. This inequality presents a limitation on the frequency. We would like to note that in the limit $H_{c1} \ll B_0 \ll H_{c2}$, where $H_{c1}$ is the lower critical field, the condition \eqref{eq:w_ll_wC} can be weakened, namely
\begin{equation}
	\omega \ll \omega_B = \frac{\Phi_0 B_0}{4\pi \lambda^2 \eta} \qquad (\omega_B \gg \omega_C).
	\label{eq:w_ll_wB}
\end{equation}
This follows directly from Equation (22) in Ref. \onlinecite{Sonin+Traito94}.

Let us turn to the case of a ferromagnetic superconductor. We assume the sample is a slab with dimensions $L_x$, $L_y$ and $L_z$, where $L_z \ll L_x,L_y$ -- see Fig. \ref{fig:Both}. The $x$-axis, parallel to the large surface of the sample, is the magnetization easy-axis. In fact, the slab geometry is not a key point for us, but the equilibrium magnetization must be parallel to one of the sample surfaces. By applying an external field we can provide that the internal field $\vec{B}_0$ is parallel to the $z$-axis. In the slab geometry the components of the demagnetizing tensor are $N_{xx} \approx 0$, $N_{yy} \approx 0$, $N_{zz} \approx 4\pi$. Then, according to Eq. \eqref{eq:F2}, if the external field is $\vec{H}_e = (-4\pi M,0,H_{ez})$, the internal field equals $\vec{B}_0 = (0,0,H_{ez})$.

Now we discuss the surface impedance of a ferromagnetic superconductor. Compared to the case of a conventional superconductor, an additional complication arises due to the presence of new degrees of freedom. These are connected with magnetization dynamics and lead to the appearance of new magnon-like modes. Such modes can be directly excited by an electromagnetic wave even in the absence of vortices,\cite{Braude+Sonin2004,Braude2006} and they may significantly influence the surface impedance. However, in our geometry the excitation of these modes can be avoided, as will be demonstrated below. 

If the frequency is not too close to the ferromagnetic resonance frequency ($\abs{\omega - \omega_F}/\omega_F \gg K^{-1}$) we can neglect the magnetostatic interaction in the Landau-Lifshitz equation when analyzing the additional magnon-like modes, as we have done in Sec. \ref{sec:general} (where the term $\vec{b}_{Mq}$ has been dropped). Then, in the limit of small dissipation, Eq. \eqref{eq:LL_lin} takes the form 
\begin{equation}
    \frac{\partial \vecm}{\partial t} = -\gamma \vec{M}_0 \times \left(\alpha \frac{\partial^2 \vecm}{\partial z^2} - K\vecm \right).
    \label{eq:LL_lin1}
\end{equation}
This yields two modes, which we label as type-3 and 4:
\begin{eqnarray}
  & \vec{m} = (\vec{z}_0 \mp i\vec{y}_0)m_{3,4} e^{ik_{3,4}z}, & \nonumber \\
  & k_3 =L^{-1} \sqrt{\frac{\omega}{\omega_F} - 1}, \qquad k_4 = iL^{-1}\sqrt{\frac{\omega}{\omega_F} + 1}, &
  \label{eq:magnons3,4}
\end{eqnarray}
where $m_3$ and $m_4$ are scalar amplitudes. Now suppose that the magnetic field $\vec{h}_e$ in the probing electromagnetic wave oscillates along the $x$-axis, i. e., along the equilibrium magnetization (see Fig. \ref{fig:Both}). We assume that inside the sample the alternating magnetic induction $\mean{\vec{b}}$, averaged over the $xy$-plane, is also parallel to the $x$-axis. It will be shown that this statement is self-consistent. Indeed, for $\mean{\vec{b}}$ parallel to $\vec{M}_0$ we see from Eqs. \eqref{eq:LL_lin} and \eqref{eq:bound_m} that $\partial{\mean{\vec{m}}}/ \partial t = 0$. This means that the magnon-like type-3 and 4 modes are not excited. In the type-1 and 2 modes $\mean{\vec{m}} = 0$, but $\mean{\vec{b}} \neq 0$. Hence, these modes differ from their analogues in non-magnetic superconductors only by the presence of the magnetic contribution to the viscosity, $\eta_M$, which is due to the Fourier-components $\vec{m}_q$ with $\vec{q} \neq 0$. Then, according to Ref. \onlinecite{Sonin+Traito94}, the internal field $\mean{\vec{b}}$ will be parallel to the probing field $\vec{h}_e$ (which follows from the London equation \eqref{eq:London}, if the deformation of the vortex lattice is taken into account). Thus, we have proved the validity of our assumption, having shown in addition that only the type-1 and 2 modes are excited.

Strictly speaking, the effective viscosity for the long-wavelength type-1 mode differs from $\eta + \eta_M$, because the vortices are not straight. However, since $\abs{k_1} \ll \lambda^{-1}$, the radius of curvature of the vortices is sufficiently large to make this difference negligible.

%This wave excites inside the sample the type-1 and 2 (and, probably, 3 and 4) modes with the magnetic induction oscillating also along the $x$-axis. Then it follows immediately from the Landau-Lifshitz equation that in these two modes the average value of $\vec{m}$ vanishes. We need to stress here that this implies only that $\vec{m}$ averages to zero on the scale of the inter-vortex distance, but its Fourier components $\vec{m}_q$ (introduced in Sec. \ref{sec:general}) with $\vec{q} \neq 0$ do not vanish, and it is they who contribute to the magnetic viscosity $\eta_M$. Strictly speaking, the effective viscosity for the long-wavelength type-1 mode differs from $\eta + \eta_M$, because the vortices are not straight. However, since $\abs{k_1} \ll \lambda^{-1}$, the radius of curvature of the vortices is sufficiently large to make this difference negligible.

%Finally, to prove that the type-3 and 4 modes are not excited, we need a boundary condition for the magnetization. If the free energy \eqref{eq:F} has no additional surface terms, the boundary condition following from the Landau-Lifshitz equation \eqref{eq:LL} is
%
%\begin{equation}
%	\frac{\partial \vec{m}}{\partial z} = 0
%	\label{eq:bound_m}
%\end{equation}
%
%on the sample surface. Since in the type-1 and 2 modes $\mean{\vec{m}} = 0$, we inevitably have to demand $m_3 = m_4 = 0$ to satisfy the boundary condition above. Hence, like in the case of a non-magnetic type-II superconductor, only two modes are excited in the sample. For this particular case Eq. \eqref{eq:Z} has been derived.

An electromagnetic wave polarized in the $y$-direction ($\vec{h}_e = h_{ey} \vec{y}_0$) requires separate treatement, which is outside the scope of this paper. Here, the magnon-like modes of type 3 and 4 must be taken into account. For a study of the surface impedance in the case $\vec{h}_e \bot \vec{M}_0$ (in a different geometry) see Ref. \onlinecite{Bespalov+2013}.

\section{Magnon excitation in SF multilayers.}
\label{sec:SFS}

In this section it is shown how our results can be extended to the case of SF multilayers with S and F being an ordinary type-II superconductor and ordinary ferromagnet, respectively. We consider structures with a sufficiently small period $d$ (see below) and with vortices oriented perpendicular to the layer surfaces -- see Fig. \ref{fig:SFS}. Then, the generalization of the results from Secs. \ref{sec:general} - \ref{sec:harmonic} is straightforward, if two points are taken into account:
\vspace{0.2cm}

\noindent (i) Since the magnetic moments now occupy only a fraction of the sample, the force $\vec{f}_M$ is reduced by a factor of $d/d_F'$, where $d_F' \leq d_F$ is the effective thickness of the ferromagnetic layer. Formally, all expressions for $\vec{f}_M$, starting with Eq. \eqref{eq:f2}, should be multiplied by $d_F'/d$. The quantities $d_F'$ and $d_F$ coincide, if the mutual influence of the superconducting and magnetic orders is negligible. However, this is not the case for cuprate/manganite superlattices. Experimental papers report giant superconductivity induced modulation of the magnetization\cite{Hoppler+2009} and the suppression of magnetic order in the manganite layer close to the SF interface.\cite{Depleted} In the latter case, $d_F'<d_F$, but both quantities are of the same order of magnitude.
\vspace{0.2cm}

\noindent (ii) Due to the fact that the structure is only partially superconducting, the in-plane London penetration depth now equals $\leff = \lambda (d/d_S)^{1/2}$ -- see Ref. \onlinecite{Klemm_book}, for example. The expression for the single vortex field
\begin{equation}
	h_{qz} \approx \frac{\Phi_0}{4\pi^2 (1+q^2 \leff^2)}
	\label{eq:hqz_leff}
\end{equation}
can be used if the period $d$ of the structure is much smaller than the characteristic in-plane length scale of the problem. To apply our results for the case of a constant driving force, we have to demand $d \ll L$, according to Sec. \ref{sec:constV}. The constraint is somewhat weaker in the case of the harmonic driving current. Indeed, for $\omega>\omega_F$ the main contribution to $\vec{f}_M$ comes from $q \approx L^{-1} \sqrt{\omega/\omega_F - 1}$, hence, the limitation on the period of the structure is
\[ d \ll L \sqrt{\frac{\omega_F}{\omega - \omega_F}}. \]
Thus, for $(\omega - \omega_F)/\omega_F \ll 1$ the thickness $d$ may be of the order of or larger than the domain wall width.

Finally, we will discuss briefly a recent paper by Torokhtii et al.,\cite{Silva+2012} where the flux-flow resistivity in Nb/PdNi/Nb trilayers has been measured. It has been reported that in the presence of the magnetic PdNi layer the flux-flow resistivity in Nb exceeds the Bardeen-Stephen estimate,\cite{Bardeen+65} as if the vortex viscosity is reduced by the interaction with magnetic moments. At first sight, this seems to contradict our prediction. However, this experiment can not be interpreted in the framework of the model used here, since the ferromagnetic alloy PdNi does not posess a well-defined magnetic anisotropy, and the magnon modes can not be characterized by a wave vector $\vec{q}$ due to the lack of translational symmetry. Moreover, the dependence of the critical temperature of Nb on the PdNi layer thickness signifies strong influence of the magnetic order on superconductivity. We suppose that the explanation of the viscosity reduction in the mentioned experiment requires a more complicated microscopic treatment.
 
\section{Conclusion}
We have calculated the magnetic moment induced force $\vec{f}_M$ acting on moving Abrikosov vortices in ferromagnetic superconductors and superconductor/ferromagnet multilayers. When the vortices are driven by a dc transport current, magnons are efficiently generated when the vortex velocity exceeds the value $\Vth = 2\omega_F L$. As a result, narrow peaks appear on the current-voltage characteristics of the superconductor, if the vortices form a regular lattice. Within a vortex lattice domain the current may be not parallel to the electric field. For a disordered vortex array a step-like feature should appear on the current-voltage characteristics. This behavior is in contrast with antiferromagnetic superconductors, where the increase of the current at the magnon generation threshold is proportional to $\sqrt{U - U_c}$, where $U$ is the voltage, and $U_c$ is some threshold value.\cite{Bulaevskii+2011} According to our estimates, the transport current required to reach the vortex velocity $\Vth$ in the U-based ferromagnetic superconductors is of the order the depairing current due to the large magnetic anisotropy of these compounds. On the other hand, in cuprate/manganite multilayers\cite{Multilayer,Hoppler+2009,Depleted} the required current is well below the depairing current, so the mentioned features may be observable on the current-voltage characteristics of such systems.

If the vortices are driven by an ac current, the interaction with magnetic moments results in the appearance of a complex magnetic contribution $\eta_M$ to the vortex viscosity. We determined this quantity for the cases of an ideal vortex lattice and a disordered vortex array. For low frequencies, $\omega \ll \omega_F$, the magnetic contribution to the vortex mass has been estimated. From the $\eta_M$ vs. magnetic field and frequency dependencies the magnon spectrum in the ferromagnetic superconductor can be extracted. Experimentally, $\eta_M$ can be determined by measuring the surface impedance of the sample in the geometry, where the equlibrium magnetization is parallel  to the oscillating external magnetic field.

\section*{Acknowledgements}
We are grateful to L. Bulaevskii for useful discussions and valuable comments. This work was supported in part by the Russian Foundation for Basic Research, European IRSES program SIMTECH (contract n.246937), the French ANR program ''electroVortex'' and LabEx ''Amadeus'' program.
\appendix

\section{}
\label{sec:Zeeman}

In this appendix we will prove that the magnetic moment induced force acting on vortices can be
written as \eqref{eq:f1.5}. We have to calculate the variation of the free energy when all vortices are shifted
by an equal vector, and the magnetization is kept fixed. To simplify the calculations we use the fact that the free energy acquires the same variation if the vortices are kept fixed, and the magnetization is shifted in the opposite direction. Then
\[ \delta F = \int \left(\frac{\delta F}{\delta \vec{A}} \delta
\vec{A} + \frac{\delta F}{\delta \vec{M}} \delta \vec{M} \right)
d^3 \vec{r}.
\]
According to the London equation $\delta F/\delta \vec{A} = 0$ the first term
in the right-hand side vanishes. Also, the terms in Eq. \eqref{eq:F}
which depend only on $\vec{M}$ (e. g., the exchange energy) are not
affected by the magnetization shift. Hence, only the term
\begin{equation}
  \delta F = - \int \vec{B} \delta \vec{M} d^3 \vec{r},
  \label{eq:deltaF1}
\end{equation}
remains, and the force acting on a vortex is
\[ (\vec{f}_M)_{x_i} = \frac{1}{N_v L_v} \int \vec{B} \frac{\partial \vec{M}}{\partial x_i} d^3 \vec{r}
= - \frac{1}{N_v L_v} \int \frac{\partial \vec{B}}{\partial x_i} \vec{M} d^3 \vec{r} \]
Presenting the magnetic field as $\vec{B} = \vec{h} + \vec{b}_M$, we have
\begin{equation}
  \vec{f}_M = \vec{f}_{M1} + \vec{f}_{M2}.
  \label{eq:f_M1}
\end{equation}
\begin{eqnarray*} 
  & (\vec{f}_{M1})_{x_i} = - \frac{1}{N_v L_v} \int \frac{\partial \vec{b}_M}{\partial x_i} \vec{M} d^3 \vec{r}, & \\
  & (\vec{f}_{M2})_{x_i} = - \frac{1}{N_v L_v} \int \frac{\partial \vec{h}}{\partial x_i} \vec{M} d^3 \vec{r}. &
\end{eqnarray*}
Note that the term $\vec{f}_{M1}$ does not depend on the vortex positions. Hence, to calculate this term
we can place the vortices anywhere in the superconductor. Let us position the vortices in an area with uniform magnetization ($\vec{M} = \mathrm{const}$). Then, $\vec{f}_{M2}$ vanishes, and $\vec{f}_{M} = \vec{f}_{M1}$.
On the other hand, in the area with homogenous magnetization $\vec{b}_M = 0$ inside the superconductor
(in the ferromagnetic superconductor this happens due to London screening, and in the SF multilayer system the field 
$\vec{b}_M$ is simply confined to the ferromagnetic layers). Hence, the magnetization has no influence on the magnetic field and
supercurrent in the vortex region, and the force $\vec{f}_{M}$ vanishes. Then, $\vec{f}_{M1} = 0$, and for any vortex positions
$\vec{f}_M = \vec{f}_{M2}$. From this follows Eq. \eqref{eq:f1.5}.

\section{}
\label{sec:Integration}

In this appendix we show how the integral in Eq. \eqref{eq:VL_disorder4} can be evaluated. We introduce the dimensionless quantities $l = L/\lambda$, $l_v = V_L/(\omega_F \lambda)$ and $\vec{g} = \lambda \vec{q}$, and direct the $g_x$-axis along $\VL$. Then $f_{My} = 0$, and
\begin{eqnarray}
	& \qquad f_{Mx} = -\frac{\gamma M \Phi_0^2 \sin^2 \theta}{4\pi \lambda^3 \omega_F} \int \frac{g_x d^2 \vec{g}}{(1+g^2)^2} \delta (1+l^2 g^2 - l_v g_x) & \nonumber \\
	& = -\frac{\gamma M \Phi_0^2 \sin^2 \theta}{4\pi \lambda^2 V_L} \int \frac{(1+l^2 g^2) \delta (1+l^2 g^2 - l_v g_x)}{(1+g^2)^2} d^2 \vec{g} & \nonumber \\
	& = -\frac{\gamma M \Phi_0^2 \sin^2 \theta}{4\pi \lambda^2 V_L} \left[ l^2 \int \frac{\delta (1+l^2 g^2 - l_v g_x)}{1+g^2} d^2 \vec{g} \right. & \nonumber \\
	& \left. + (1-l^2) \int \frac{\delta (1+l^2 g^2 - l_v g_x)}{(1+g^2)^2} d^2 \vec{g} \right]. &
	\label{eq:Integral1}
\end{eqnarray}
Now we make a coordinate shift, redesignating $g_x - l_v/(2l^2)$ by $g_x$:
\begin{eqnarray}
	& f_{Mx} = -\frac{\gamma M \Phi_0^2 \sin^2 \theta}{4\pi \lambda^2 V_L} \left\{ \int \frac{\delta(g^2 - g_0^2) d^2 \vec{g}}{1 + g_y^2 + \left( g_x + \frac{l_v}{2l^2} \right)^2} \right. & \nonumber \\
	& \left. + (l^{-2} - 1) \int \frac{\delta(g^2 - g_0^2) d^2 \vec{g}}{\left[1 + g_y^2 + \left( g_x + \frac{l_v}{2l^2} \right)^2 \right]^2} \right\}, &
	\label{eq:Integral2}
\end{eqnarray}
where
\[ g_0^2 = l^{-2} \left( \frac{l_v^2}{4l^2} - 1 \right). \]
Further we assume that $V_L>\Vth$, so that $g_0^2 >0$ (at $V_L<\Vth$ $\vec{f}_M = 0$). Integration over the modulus of $\vec{g}$ is now straightforward. Then
\begin{eqnarray}
	& f_{Mx} = -\frac{\gamma M \Phi_0^2 \sin^2 \theta}{8\pi \lambda^2 V_L} \left[ \int_0^{2\pi} \frac{d\varphi}{1+g_0^2 + \frac{l_v^2}{4l^4} + \frac{l_v}{l^2} g_0 \cos \varphi} \right. & \nonumber \\
	& \left. + \int_0^{2\pi} \frac{(l^{-2} - 1) d\varphi}{\left( 1+g_0^2 + \frac{l_v^2}{4l^4} + \frac{l_v}{l^2} g_0 \cos \varphi \right)^2} \right], &
\end{eqnarray}
where $\varphi$ is the polar angle in the $\vec{g}$-plane. Integration can be completed using standard methods or a table of integrals. The result is
\begin{equation}
	f_{Mx} = -\frac{\gamma M \Phi_0^2 \sin^2 \theta}{8 \lambda^2 V_L} (l^{-2} + 1) \frac{l_v^2}{l^4} \left[ (1-l^{-2})^2 + \frac{l_v^2}{l^4} \right]^{-3/2}.
	\label{eq:Integral3}
\end{equation}
If we return to dimensional variables and recall that $L \ll \lambda$, we obtain Eq. \eqref{eq:VL_disorder}.

\end{document}